\documentclass[aps, prb, twocolumn, footinbib]{revtex4-1}

\usepackage{enumerate}
\usepackage{comment}
\usepackage{amsmath}
\usepackage{amssymb}
\usepackage{color}
\usepackage[dvipsnames]{xcolor}
\usepackage{tikz}
\usepackage{amsthm}
\usepackage{graphicx}
\usepackage{bbold}
\usepackage{empheq}
\usepackage{subfigure}
\usepackage{xfrac}
\newcommand{\be}{\begin{equation}}
\newcommand{\ee}{\end{equation}}
\newcommand{\bea}{\begin{eqnarray}}
\newcommand{\eea}{\end{eqnarray}}
\newcommand{\beb}{\begin{eqnarray*}}
\newcommand{\eeb}{\end{eqnarray*}}

\newcommand{\LD}{\langle}
\newcommand{\RD}{\rangle}
\newcommand{\Hh}{\text{H}}

\newcommand{\eqn}{Eq.~}
\newcommand{\eqns}{Eqs.~}
\newcommand{\fig}{Fig.~}
\newcommand{\figs}{Figs.~}

\newcommand{\SPstate}{\Psi }
\newcommand{\SPWF}{\varphi }

\newcommand{\TP}{two-particle }
\newcommand{\SP}{single-particle }

\newcommand{\Hzero}{\hat{H}^0}
\newcommand{\HC}{\hat{V}}
\newcommand{\HSR}{\hat{U}}
\newcommand{\TPorbWF}{\Phi}
\newcommand{\TPEn}{\mathfrak{E}}

\newcommand{\BLG}{BLG}

\newcommand{\RN}[1]{%
  \textup{\uppercase\expandafter{\romannumeral#1}}%
}

\begin{document}

\title{Quartet states in two-electron quantum dots in bilayer graphene}

\author{Angelika Knothe$^{1}$}
\author{Vladimir Fal'ko$^{1,2,3}$}
\affiliation{$^1$National Graphene Institute, University of Manchester, Manchester M13 9PL, United Kingdom}
\affiliation{$^2$Department of Physics and Astronomy, University of Manchester, Oxford Road, Manchester, M13 9PL, United Kingdom}
\affiliation{$^3$Henry Royce Institute for Advanced Materials, University of Manchester, Manchester, M13 9PL, United Kingdom}
\date{\today}
 
\begin{abstract}
Trapping electrons in quantum dots and   controlling their collective quantum states is crucial for converting semiconductor structures into bits of quantum information processing. Here, we study single- and two-particle states in quantum dots formed in gapped bilayer graphene (BLG), where the electron's valley states appear in pair with their spin quantum number and we analyse spin- and valley-singlet and triplet states for various BLG and dot parameters, as well as two-particle interaction strength and external magnetic field.
\end{abstract}
 
\maketitle

Few-electron quantum dots (QD) were extensively investigated in various semiconductors (Si \cite{Delley1993, Wang1994}, GaAs/AlGaAs heterostructures \cite{Cibert1986, Plaut1991, Brunner1992}, varieties of type III - IV semiconductors \cite{Sikorski1989, Kash1990, Petroff1994}), where the understanding of the QD's ground state (GS), excitations, and addition spectra led to the suggestions \cite{Loss1998} for their use as solid state qubits\cite{Zhang2018, Wang2018, Tarucha2016, Cogan2018, Hanson2007}.  
 The extensive studies of few-electron states in QDs made of conventional gap-full semiconductors have already resulted in a complete understanding of their GS and excitation properties. Here, we study  two-electron states (2ES) in QDs based on a semimetal with a  non-trivial multi-valley band structure, namely, bilayer graphene (BLG). Due to the features of the BLG electronic spectrum, the theoretical description of 2ES in BLG QDs  lacks the elegant exact solutions developed for two electrons with parabolic dispersion \cite{Pfannkuche1991, Maksym1993}, and, here, we develop a combined analytical and computational approach to describe their low-energy spectra.

\begin{figure}[t!]
 \centering
\includegraphics[width=1\linewidth]{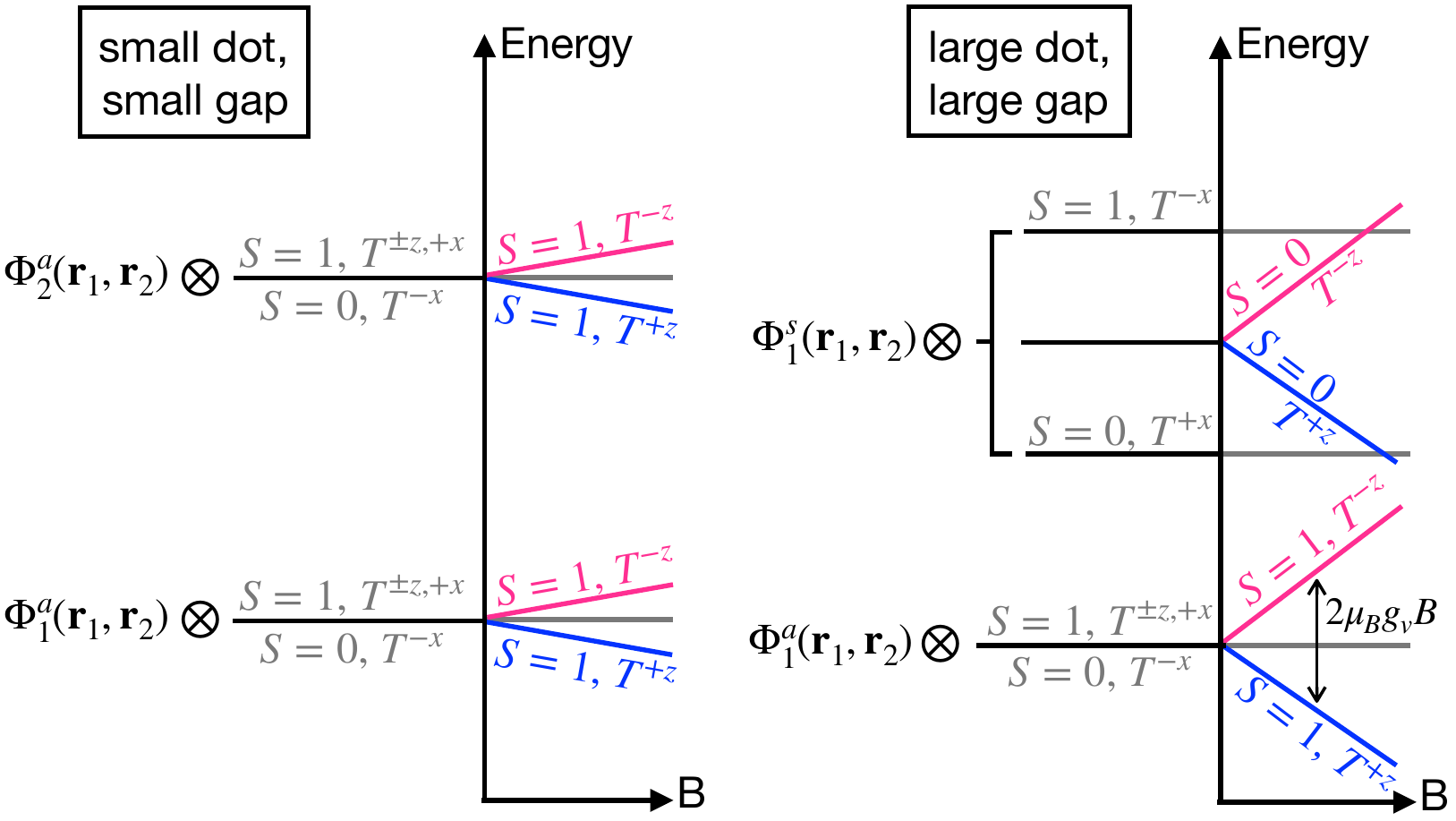}
\caption{Level schemes for two strongly interacting electrons confined in a QD in BLG with  a weak  gap (left), or a strong gap (right): a \TP state where the orbital part of of the wave function is antisymmetric, $\TPorbWF^a$, minimizes the charging energy and  entails the GS. Higher-energy states can be orbitally symmetric or antisymmetric. Short-range interactions determine splittings between spin singlets ($S=0$) and triplets ($S=1$), and valley singlets ($T^{-x}$) and triplets   ($T^{\pm z, +x}$) . Valley polarized states are split linearly in  a weak magnetic field\cite{Note1} according to their valley g-factor, $g_v$.}
\label{fig:QD_Levelstructure}
\end{figure}

 To create QDs in BLG, one needs to open an interlayer asymmetry gap \cite{McCann2006} in this, otherwise, gapless semiconductor and, then, to confine electrons using a combination of top and bottom gates \cite{Eich2018c, Overweg2018, Kraft2018, Banszerus2019, Banszerus2019a}. Upon opening a gap, the spectrum of electrons of both, the conduction, and the valence band edges acquire a strongly non-parabolic  dispersion \cite{Varlet2014, Varlet2015, Knothe2018}, which even has  a slightly inverted form featuring three shallow minivalleys  around both, $K^+$ and $K^-$, valleys. This comes on top of graphene's valley  (T) and spin (S) quantum numbers. The flat band edge  promotes the formation of different spin and valley multiplets for the  low-energy  \TP QD states,  illustrated in \fig\ref{fig:QD_Levelstructure}. Typically, we find that the GS is a multiplet that includes simultaneous spin and valley singlets and triplets, prescribed by the electron-electron exchange interaction. Depending on the dot size and the interlayer asymmetry gap, the first excited state may belong to a different spin/valley multiplet, with fine energy splittings  determined by inter- and intra-valley interactions between electrons. Additionally, one can use an external magnetic field to tune the valley splitting \cite{Knothe2018, Lee2019, Moulsdale2020} due to the valley-dependent topological magnetic moment of the electrons\footnote{The valley g-factor assumes values of $g_v\sim100$ in BLG, and therefore exceeds the spin g-factor, $g\sim2$, by orders of magnitude. Any spin splitting is hence neglected in our discussion (c.f. Refs. [26,27])}.

Below, we describe  interacting electrons in a BLG QD using a Hamiltonian,
\begin{equation}
\hat{\Hh}=\Hzero+ \HC + \HSR.
\label{eqn:Htotal}
\end{equation}
Here, $\Hzero$, is a single-particle part (specified below) taking into account the BLG dispersion and  a smooth electrostatic confinement. The operator  $ \HC$,  acts on the orbital part,  $\TPorbWF_{\alpha} (\mathbf{r}_1,\mathbf{r}_2)$,   of the \TP wave function,
\begin{equation}
 V_{\alpha,\beta}\! =\! \iint \! d\mathbf{r}_1d\mathbf{r}_2\;\TPorbWF^*_{ { \alpha}}(\mathbf{r}_1,\mathbf{r}_2)  V(\mathbf{r}_1-\mathbf{r}_2) \TPorbWF_{ { \beta}}(\mathbf{r}_1,\mathbf{r}_2),
\label{eqn:int}
\end{equation}
and takes into account  the 2D screened Coulomb interaction in a weakly gapped BLG \cite{Cheianov2012}. Its  Fourier representation, $V(\mathbf{q})=\frac{e^2}{4\pi\epsilon_0\epsilon}\frac{2\pi}{q(1+q R_{\star})}$ (where  $\epsilon_0$ is the vacuum permittivity and $\epsilon $ is the dielectric constant of the encapsulating substrate material, such as hBN), takes into account  the polarisability of gapped BLG\cite{Cheianov2012}, with  $\kappa^2 =2m e^2/(4\pi\epsilon_0\epsilon \hbar \sqrt{\Delta} )^2  $ ($m$ being the effective mass). It results in a Keldysh-like potential \cite{Rytova2018, Keldysh1979, Keldysh1979a}, $V(r< R_{\star})\propto \frac{e^2}{\epsilon R_{\star}} \ln (r/R_{\star})$, and  $V(r\gg R_{\star})\propto \frac{e^2}{\epsilon r}$ (with $R_{\star}=  \sqrt{32} \hbar\kappa /\sqrt{\Delta}$). 
The last term in \eqn\eqref{eqn:Htotal} takes into account short-range (lattice-constant-scale) electron-electron interactions\cite{Lemonik2010, Lemonik2012, Aleiner2007},
\begin{align}
\nonumber&\HSR=\hat W\otimes\begin{pmatrix}
    g_{zz} & 0 & 0&0\\
 0 &    g_{zz}&4g_{\perp} &0 \\
0& 4g_{\perp}&     g_{zz}& 0\\
0& 0& 0&   g_{zz}
\end{pmatrix}, \\
\label{eqn:HSR}
&W_{\alpha,\beta}=\frac{1}{2}\int d\mathbf{r}\, \TPorbWF^*_{\alpha}(\mathbf{r},\mathbf{r}) \TPorbWF_{\beta}(\mathbf{r},\mathbf{r}),
\end{align}
which is written  in the basis of valley Bloch states, $\{ {K_1^+K_2^+},  {K_1^+K_2^-},  {K_1^-K_2^+},  {K_1^-K_2^-}\}$,   of the two electrons ($\xi=\pm1$ is the valley index). The coupling parameters $g_{zz}$ (intra-valley) and $g_{\perp}$ (inter-valley short-range interaction) and the role they play in splitting the low-energy multiplets will be discussed later in the text.

\begin{figure}[t!]
 \centering
\includegraphics[width=1\linewidth]{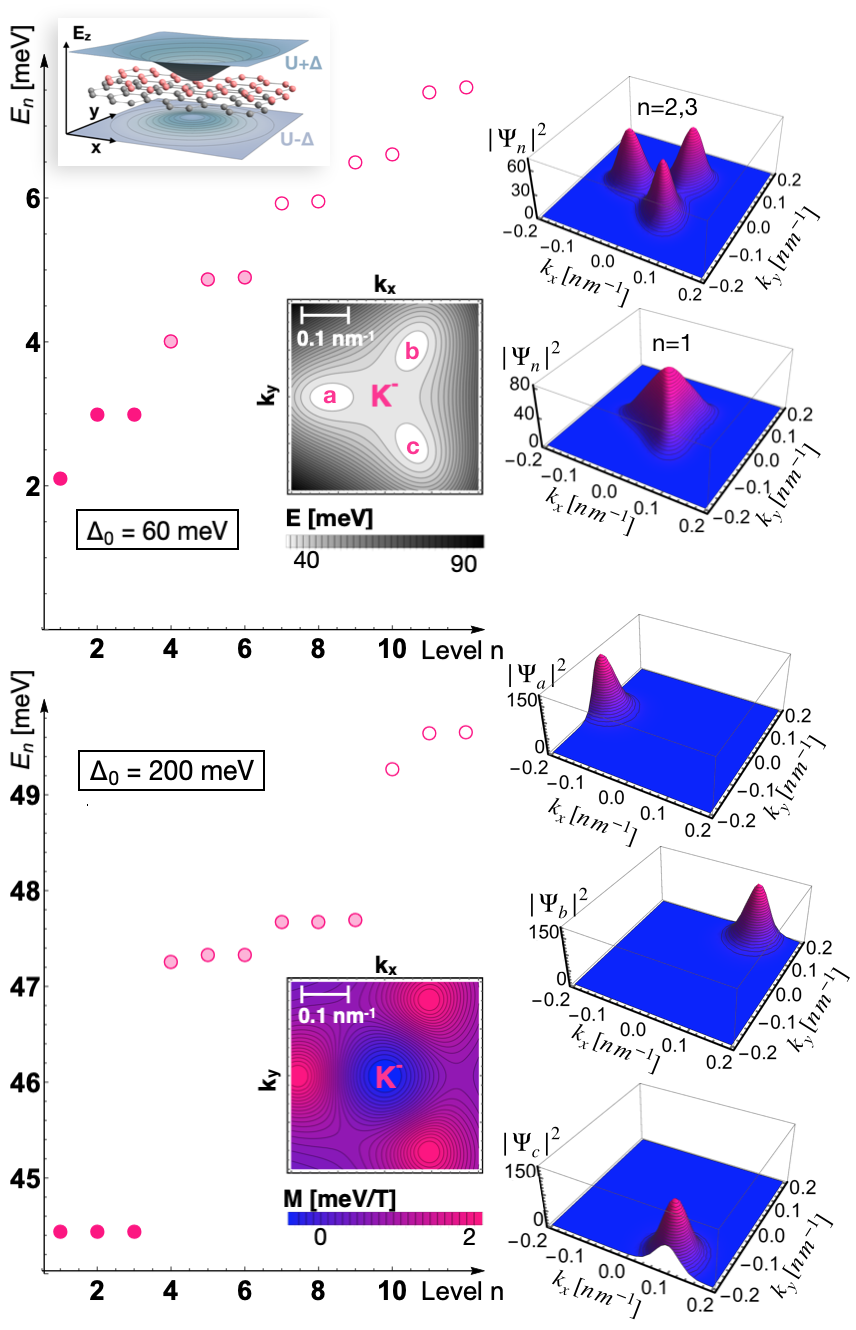}
\caption{Single-particle energy spectra for $U_0=-20$ meV, $L=80$ nm, and a small gap ($\Delta_0=60$ meV) or a large gap ($\Delta_0=200$ meV), demonstrating the change of multiplicity as the GS develops from a non-degenerate state to a threefold-degenerate state. Shadings of the dots indicate which levels are populated by a two-electron state: For  no or weak interactions, only the lowest states are populated (dark circles). For stronger interaction, states are mixed and higher energy states are involved (light circles). The higher lying states remain unpopulated (empty circles).  To the right, we show the probability distribution, $|\SPstate_n|^2$, in valley $K^-$ for the lowest levels $n=1,2,3$ (small gap), or the states in the  minivalleys $n=a,b,c$ (large gap). Insets: Electrostatic potential realizing the confinement; dispersion, E, of the first conduction band of gapped BLG with a gap of $\Delta=70$ meV; corresponding orbital magnetic moment, $M$, around the $K^-$-valley.}
\label{fig:SPproperties}
\end{figure}

For a  BLG QD, formed with the help of electrostatic  split gates\cite{Eich2018a, Overweg2018, Overweg2018a} , we employ a circularly symmetric potential, $U(\mathbf{r})$, and a gap profile, $\Delta(\mathbf{r})$, which enter in the single-electron four-band Hamiltonian\cite{McCann2007, McCann2013},
\begin{align}
 \nonumber \Hzero_{\pm}\!=\! \!&
\setlength{\arraycolsep}{+2pt} \begin{pmatrix} 
 U \mp\frac{1}{2}\Delta  & \pm v_3\pi & 0 &\pm v \pi^{\dagger}\\
\pm v_3 \pi^{\dagger}&  U \pm\frac{1}{2}\Delta  & \pm v\pi &0\\
 0 & \pm v\pi^{\dagger} &   U \pm\frac{1}{2}\Delta  &   \gamma_1\\
\pm v\pi & 0 &   \gamma_1 &  U \mp\frac{1}{2}\Delta 
\end{pmatrix}\!\!,\\
& \hskip5pt U(\mathbf{r})=\frac{U_0}{\cosh{\frac{r}{L}}}, \hskip10pt \Delta(\mathbf{r})=\Delta_0-\frac{0.3\Delta_0}{\cosh{\frac{r}{L}}}.
\label{eqn:H}
\end{align}
with $\mathbf{r}=(x,y)$, $r=|\mathbf{r}|$, momenta $\pi=p_x+ip_y,\,  \pi^{\dagger}=p_x-ip_y$ 
(where $\mathbf{p}=-i\hbar\nabla$), velocities $v=1.02*10^6 \text{ m/s}$ and  $v_3\approx0.12 v$, and energy $\gamma_1\approx0.38\text{ eV}$. This Hamiltonian  is written for the Bloch function components $\psi_{K^+}=(\psi_{A},\psi_{B^{\prime}},\psi_{A^{\prime}},\psi_{B})$ in valley $K^+$, and $\psi_{K^-}=(\psi_{B^{\prime}},\psi_{A},\psi_{B},\psi_{A^{\prime}})$ in valley $K^-$,  with electron's amplitudes on the BLG sublattices  $A$ and $B$ in the top, and  $A^{\prime}$ and $B^{\prime}$ in the bottom layer. 
In the absence of confinement, \eqn\eqref{eqn:H} describes the low energy trigonally warped bands \cite{Varlet2014, Varlet2015, Knothe2018} featuring three minivalleys around each $K$ point (\fig\ref{fig:SPproperties} inset).

To study the 2ES properties of the QD, we proceed as follows. First, we numerically diagonalize  the Hamiltonian in \eqn\eqref{eqn:H} in a  basis of  localized states (SI section S1) and obtain the \SP spectrum of the QD, $\SPstate_n(\mathbf{r})$. Using those states, we construct a basis of symmetrized (s) or antisymmetrized (a)  \TP orbital wave functions,
\begin{align}
\phi^{s/a}_{ ij}(\mathbf{r}_1,\mathbf{r}_2)=\frac{1}{\sqrt{2}} \big[\SPstate_{i}(\mathbf{r}_1)\SPstate_{j}(\mathbf{r}_2) \pm \SPstate_{i}(\mathbf{r}_2)\SPstate_{j}(\mathbf{r}_1)\big],
\label{eqn:SPsa}
\end{align}
and compute the matrix elements of the interaction operators $\HC$ and $\HSR$ in \eqn\eqref{eqn:Htotal}. Diagonalizing the resulting matrix yields the QD 2ES energy spectrum, $\TPEn$, and wave functions, $\TPorbWF^{s/a}(\mathbf{r}_1,\mathbf{r}_2)=\sum_{\alpha} \lambda^{s/a}_{\alpha}\phi^{s/a}_{ \alpha}(\mathbf{r}_1,\mathbf{r}_2)$, $\alpha=ij$ (SI section S5). Note that for the symmetric states, we sum over $\alpha \in\{11,12,\dots\}$, whereas for the antisymmetric states indices run over   $\alpha \in\{12,13,\dots\}$ (excluding nn combinations).
 
\begin{figure}[t!]
 \centering
\includegraphics[width=1\linewidth]{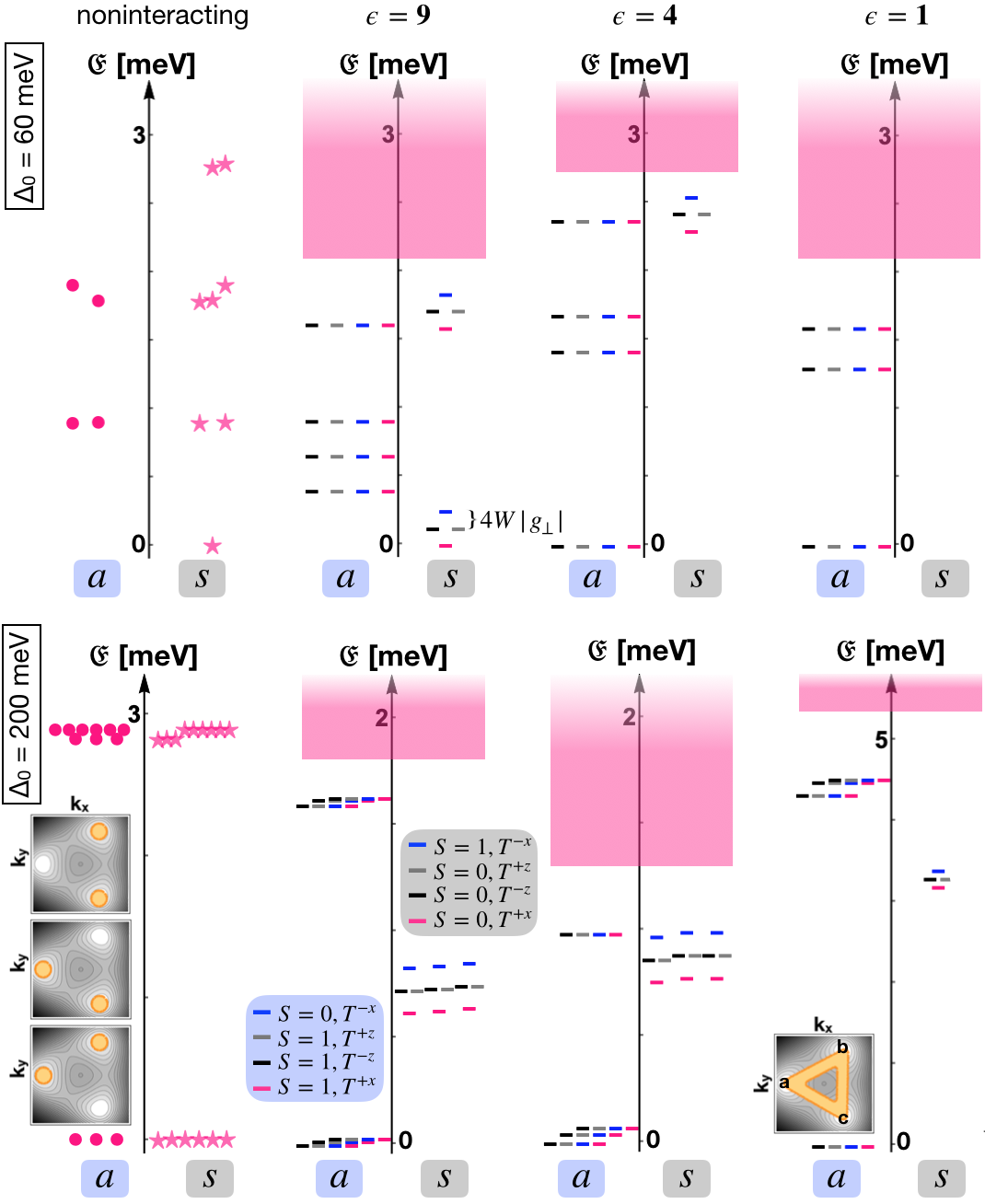}
\caption{Two-particle spectra for interacting electrons depending on the dielectric constant, $\epsilon$, of the encapsulating medium in QDs in BLG with a small gap (top row,  $\Delta_0$=60 meV) and a large gap (bottom row,  $\Delta_0$=200 meV) computed in the subspace of the $N=6$ or $N=9$ lowest \SP states $\SPstate_n(\mathbf{r})$. For the noninteracting case, magenta dots (stars) represent the energies of antisymmetric (symmetric) orbital states, each four-fold degenerate in the spin and valley degree of freedom. For non-zero interactions, we identify the  spin/valley configurations of the multiplets by lines of different color and indicate the onset of a continuum  by a magenta bar. The case $\epsilon = 4$ (BLG encapsulated by hBN\cite{Levinshtein2001, Rytova2018}) corresponds to \fig\ref{fig:QD_Levelstructure}. Insets portray mini-valley-incoherent (noninteracting) or mini-valley-coherent (strongly interacting, $\epsilon=1$) 2ESs.}
\label{fig:QD_MPSpactraBoth}
\end{figure}

In  \fig\ref{fig:SPproperties}, we illustrate \SP spectra, $E_n$, for a dot with $U_0=-20$ meV, $L=80$ nm and different values of $\Delta_0$. Alongside this, we plot the probability distribution, $|\SPstate_n(\mathbf{k})|^2$, (in valley $K^-$) for the lowest three levels. The trigonally warped mini-valley structure determines  $C_3$ rotational symmetry of the system, even for a circularily symmetric confinement potential. This symmetry breaking lifts the degeneracies of the usual Fock-Darwin levels\cite{Fock1928, Darwin1931}, commonly used for describing the discrete electronic states of a QD in 2D electron gases under circular harmonic confinement (see SI section S2 and S3).  
For smaller dots in BLG with a smaller $\Delta_0$ and stronger confinement,  the Fourier transform of the wave function is squeezed towards the center of the $K$-valleys in momentum space (top left panel of \fig\ref{fig:SPproperties}). This suppresses the minivalley effect, leading to a non-degenerate GS, $\SPstate_1$,  isolated from the rest of the spectrum by  several meV. Higher states, $\SPstate_n$ with $n>1$, come in approximate doublets. Upon increasing $\Delta_0$, the \SP level structure evolves into clearly identifiable  mini-valley triplets, as shown in \fig\ref{fig:SPproperties} for $\Delta_0=200$ meV. For a  large enough gap,  the minivalleys are sufficiently developed and separated far enough in momentum space to represent a good quantum number.  In this regime, all the \SP levels are  threefold degenerate  (three minivalleys in each valley, the corresponding \SP wave functions are shown in \fig\ref{fig:SPproperties} in momentum space and in \fig \ref{fig:QD_MPSpactraBoth} in real space). For this reason, it is beneficial to identify the basis of minivalleys, which we label by $a,b,c$ (and use below for the interpretation of the 2ES),
\begin{align}
\nonumber\SPstate_a &= (\SPstate_1+\SPstate_2+\SPstate_3)/{\sqrt{3}},\\
\nonumber \SPstate_b &=  (\SPstate_1+e^{-{i\frac{2\pi}{3}}}\SPstate_2+e^{{i\frac{2\pi}{3}}}\SPstate_3)/{\sqrt{3}},\\
\SPstate_c &= (\SPstate_1+e^{{i\frac{2\pi}{3}}}\SPstate_2+e^{{-i\frac{2\pi}{3}}}\SPstate_3)/{\sqrt{3}}.
\label{eqn:MVBasis}
\end{align}

To mention, in SI section S4, we describe  selection rules for optical transitions between these for both types of spectra shown in \fig\ref{fig:SPproperties}. Also, the states of gapped BLG carry a non-zero orbital magnetic moment, $M_z(\pm K)=\pm \mu_B g_v(\mathbf{k})  $, \cite{Xiao2010, Chang1996a}, which has opposite signs in the opposite valleys. It is convenient to characterise this quantity using a valley g-factor, $g_v$, that is small for the states exactly at the centre of the $K^{\pm}$ valleys and reaches a magnitude of $g_v\sim 10^2$ at the mini-valley dispersion minima\cite{Knothe2018, Moulsdale2020}, (inset to \fig\ref{fig:SPproperties}). The valley dependence of $M_z(k)$ leads to the valley splitting, $ 2 B_z \mu_B$ in an external magnetic field\cite{Note1}, which we take into account at the end of this analysis in relation to the fine structure of the 2ESs.  An orbital magnetic moment implies some angular momentum, whose coupling to the orbital motion of an electron results in a small splitting of the higher QD levels, $E_2$ and $E_3$ (see SI section S2).

To find the spectra of two electron states in a dot we solve the eigenvalue equation, 
\begin{widetext}
\begin{align}
\nonumber &\hat{\Hh}  \TPorbWF_{ }=\TPEn_{ }  \TPorbWF_{ }\Leftrightarrow \sum_{ \beta=kl} \big[ E_i \delta_{ik}+ E_j \delta_{jl} + V^{s/a}_{\alpha,\beta}  +\delta\TPEn_{\alpha,\beta}^{s/a}(\tau) \big]  \lambda^{s/a}_{\beta}=\TPEn^{s/a}_{} \lambda^{s/a}_{\alpha},\; \alpha=ij,\\
\nonumber&V^{s/a}_{\alpha,\beta}=V_{ik,jl}\pm V_{il,jk}, \; V_{ij,kl}=\iint d\mathbf{r}_1d\mathbf{r}_2\; \SPstate_{i}^*(\mathbf{r}_1)\SPstate_{j}(\mathbf{r}_1)V(\mathbf{r}_1-\mathbf{r}_2) \SPstate_{k}^*(\mathbf{r}_2)\SPstate_{l}(\mathbf{r}_2) , \\
\nonumber& \delta\TPEn_{\alpha,\beta}^{s}(T^{\pm z})= W_{\alpha,\beta}\, g_{zz},\;\;  \delta\TPEn^s_{\alpha,\beta}(T^{\pm x})=W_{\alpha,\beta}\, (g_{zz}\pm g_{\perp}), \; \; \delta\TPEn^a \equiv0 ,\\
& W_{\alpha,\beta}= \int d\mathbf{r}  \SPstate_{i}^*(\mathbf{r}) \SPstate_{j}^*(\mathbf{r}) \SPstate_{k}(\mathbf{r}) \SPstate_{l}(\mathbf{r}).
\label{eqn:EVeqnS}
\end{align}
\end{widetext}

In \eqn\eqref{eqn:EVeqnS}, we distinguish between the symmetric and antisymmetric combinations of \SP orbitals, as introduced in \eqn\eqref{eqn:SPsa}, which is also related to the  structure of multiplets, $\tau$, involving valley degrees of freedom. "Valley triplet" states comprise the polarized states,   $\tau=T^{\pm z}$ ($K_1^+K_2^+$ and $K_1^- K_2^-$),  and the coherent combination, $\tau=T^{+x}$ ($[K_1^+K_2^- + K_1^-K_2^+]/\sqrt{2}$), and they are complemented by the  "valley singlet", $\tau=T^{-x}$ ($[K_1^+K_2^- - K_1^-K_2^+]/\sqrt{2}$). These valley multiplets  come on top of the conventional classification of spin-singlet and -triplet states\cite{Note1}.

Depending on the  interaction strengths, various combinations of valley-/spin-singlets and -triplets appear as  the low-energy eigenstates of the Hamiltonian in \eqn\eqref{eqn:EVeqnS}. For sufficiently strong long-range interaction, $\HC$, the GS will  be an antisymmetric combination of orbitals, $\TPorbWF^{a}$, as favored by the exchange interaction. For weak long-range interactions (compared to the gaps in the noninteracting 2ES spectrum), orbitally symmetric GSs may occur,  $\TPorbWF^{s}$, for which different valley states are shifted by  $W_{\alpha,\beta}\,g_{zz}$  and split proportionally to $W_{\alpha,\beta}\,g_{\perp}$,  according to  \eqn\eqref{eqn:EVeqnS}.  

To compare  different interaction energies in \eqn\eqref{eqn:EVeqnS}, we  estimate  the valley splittings of the orbitally symmetric states, $\TPorbWF^{s}(\mathbf{r}_1,\mathbf{r}_2)$.  From the lowest  \SP wave functions,  $\SPstate_{1}$,  for  $L=80$ nm and a gap of $\Delta_0=60$ meV we calculate  $W_{11,11} =4.3*10^{-4}$ nm$^{-2}$.  For a rough estimate of numerical values for the couplings $g_{zz, \perp}$, we use the $p_z$-orbitals in the BLG Bloch states (SI section S6), and obtain $g_{zz}=\sfrac{4.04}{\epsilon_c}$ eVnm$^2$, $g_{\perp}=\sfrac{-0.191}{\epsilon_c}$ eVnm$^2$, in accordance with previous estimations of the microscopic coupling parameters in monolayer and bilayer graphene\cite{Kharitonov2012a, Kharitonov2012b}, which likewise suggested $g_{zz}>0, g_{\perp}<0$ and $g_{zz} > |g_{\perp}|$. Here, $\epsilon_c\approx2.65$, is  the dielectric susceptibility of BLG\cite{Slizovskiy2019}. We estimate a shift of the orbitally symmetric states of $W_{11,11} \; g_{zz}=0.66$ meV, and a  valley splitting of $4|W_{11,11} \;g_{\perp}|= 0.14$ meV. For $\Delta_0=200$ meV we obtain $W_{11,11}=3.7*10^{-4}$ nm$^{-2}$, entailing $W_{11,11} \; g_{zz}=0.56$ meV and $4|Wg_{\perp}|=0.11$ meV\footnote{Note that the values and the sign of the short-range couplings could be affected further by a more realistic shape of the atomic orbitals or by renormalization$^{34}$. A positive value of $g_{\perp}$ would invert the order of  states within the split valley multiplets}. Below, we will  drop the indices for brevity,  implying   $W=W_{11,11}$.

Combining the latter estimates with the numerical evaluation of the Coulomb matrix elements in \eqn\eqref{eqn:EVeqnS}, we calculate the spectra for the  2ES, such as those illustrated in \fig \ref{fig:QD_MPSpactraBoth} for several values of the dielectric constant $\epsilon$ of the encapsulated medium and for  $\Delta_0=60$ meV and $\Delta_0=200$ meV (energies of the states are listed in SI section S7). We distinguish between the spectra of orbitally antisymmetric and orbitally symmetric  states, and identify the splittings of valley multiplets. %Alongside the spectra in \fig \ref{fig:QD_MPSpactraBoth}, we display the real space probability distributions of the lowest \SP states used to construct the 2ESs.
 
In  the absence of interactions ($\epsilon\rightarrow\infty$), 2ESs are exact combinations of the  \SP states as written in \eqn\eqref{eqn:SPsa}. In the case of a small gap,  $\Delta_0=60$ meV, and $\epsilon = \infty$, the \TP GS is  the singly-degenerate state $\TPorbWF^{s}_{11}$  (upper left panel  in \fig\ref{fig:QD_MPSpactraBoth}), since the \SP GS,  $\SPstate_1$, is non-degenerate and therefore $\TPEn_{11}^s<\TPEn_{12}^a=\TPEn_{13}^a$. This orbital composition of the GS is stable  against a weak   interaction ($\epsilon=9$  in \fig\ref{fig:QD_MPSpactraBoth}) where the electron replusion is not sufficient to alter  the orbital wave function. Hence, in the weakly interacting regime,   the first orbital symmetric state, $\TPorbWF^{s}_{11}$, remains to be the basis for the GS and the corresponding spin and valley levels are shifted and split according to the short-range interaction part, $\delta\TPEn^{s}$, in \eqn\eqref{eqn:EVeqnS}. As the overall wavefunction should be antisymmetric, the  resulting GS  is a spin-singlet ($S=0$) and a valley  triplet ($T^{+x}$). This form of the QD GS resembles a singlet Cooper pair as in the superconducting phase discussed previously in BLG as a consequence of short-range interactions\cite{Lemonik2012}. 

Conversely, if the gap is large enough for the  gapped BLG mini-valleys to survive size quantization ($\Delta_0=200$ in \fig\ref{fig:QD_MPSpactraBoth}), the noninteracting \TP GS acquires an additional degeneracy. Orbitally symmetric and antisymmetric states can occur at equal energy, $\TPEn_{11}^s=\TPEn_{12}^s=\TPEn_{13}^s=\TPEn_{22}^s=\TPEn_{23}^s=\TPEn_{33}^s=\TPEn_{12}^a=\TPEn_{13}^a=\TPEn_{23}$, due to the three mini-valley states of the \SP GSs,  $\SPstate_1$, $\SPstate_2$, and $\SPstate_3$ (bottom left panel  in \fig\ref{fig:QD_MPSpactraBoth}). This latter degeneracy is  conveniently exploited using the basis of minivalleys states $\SPstate_a, \SPstate_b, \SPstate_c$ (see \eqn\eqref{eqn:MVBasis}).  Then, combinations of the orbitally asymmetric 2ESs, $\TPorbWF^{a}_{ab}$, $\TPorbWF^{a}_{ac}$, and $\TPorbWF^{a}_{bc}$, where each electron sits in its own mini-valley  (shown as insets to the bottom left panel of  \fig\ref{fig:QD_MPSpactraBoth}), determine the three-fold degenerate GS of the system at zero or weak interactions:
 \begin{align}
\nonumber\TPorbWF^{a}_{12}&= (\TPorbWF^{a}_{ab}+e^{i\frac{\pi}{3}} \TPorbWF^{a}_{ac}+e^{i\frac{2\pi}{3}} \TPorbWF^{a}_{bc})/\sqrt{3},\\
 \nonumber \TPorbWF^{a}_{13} &=  (\TPorbWF^{a}_{ab}+e^{-i\frac{\pi}{3}} \TPorbWF^{a}_{ac}+e^{-i\frac{2\pi}{3}} \TPorbWF^{a}_{bc})/\sqrt{3}, \\
\TPorbWF^{a}_{23}&=   (\TPorbWF^{a}_{ab}-\TPorbWF^{a}_{ac}+\TPorbWF^{a}_{bc} )/\sqrt{3}.
\label{eqn:Phiab}
\end{align}
 In the strongly gapped case,  already for marginally weak long-range interactions (as for $\epsilon=9$ in \fig\ref{fig:QD_MPSpactraBoth}),  the orbitally antisymmetric configurations are  favored energetically over the symmetric ones, (as there is no  advantage in kinetic energy for orbitally symmetric states compared to the orbitally asymmetric ones) and the splitting within the lowest-energy triplet increases with increasing interaction strength.

 In the  limit when long-range repulsion is a dominant factor ($\epsilon\rightarrow1$), the \TP GS is  orbitally antisymmetric  for all system parameters. Also, in the latter case,   the 2ESs are more extended compared to the size of the lowest \SP orbitals, and, due to the more significant exchange gaps,  higher orbital \SP states are involved in forming the 2ES. Therefore, the 2ES in the case $\Delta_0=200$, $\epsilon=1$ in \fig\ref{fig:QD_MPSpactraBoth}, is the antisymmetric combination of $ (\SPstate_a+\SPstate_b+\SPstate_c) (\tilde{\SPstate}_a +\tilde{\SPstate}_b +\tilde{\SPstate}_c  )/3 $, where $\SPstate_{a}$,$\SPstate_{b}$,$\SPstate_{c}$ denote the mini-valley states of the lowest \SP triplet  as in \eqn\eqref{eqn:MVBasis}, and   $\tilde{\SPstate}_{a}$, $\tilde{\SPstate}_{b}$, $\tilde{\SPstate}_{c}$ the corresponding mini-valley states in the third triplet (see  SI section S2 for  details about  these wave functions). This 2ES state is an inter-mini-valley coherent state (see inset to  \fig\ref{fig:QD_MPSpactraBoth}), where all the mini-valleys of the lowest, and the third triplet are occupied simultaneously (as opposed to the incoherent population of one electron per mini-valley in \eqn\eqref{eqn:Phiab}). For the first excited state there is a competition between the second orbitally antisymmetric state  (as in the case $\Delta_0=60$, $\epsilon=1$ in  \fig\ref{fig:QD_MPSpactraBoth}), or the first orbitally symmetric state (which is the first excited state for $\Delta_0=200$, $\epsilon=1$, in  \fig\ref{fig:QD_MPSpactraBoth}). Implied that, depending on the size of the dot and the magnitude of the gap, the first excited states can  be four-fold spin and valley degenerate (if the orbital part is antisymmetric), or manifest spin and valley-split levels (if the orbital wave function is symmetric).

The GS multiplets can be split in a magnetic field, $B$. The valley g-factor, typically $g_v\gg1$\cite{Knothe2018, Overweg2018a, Lee2019}, determines the valley splitting of states by $B$\cite{Note1}. 
For example, the \SP mini-valley triplet states in the strongly gapped BLG in \fig\ref{fig:SPproperties}, remain degenerate in the regime of weak $B$. For the 2ESs, the valley polarized states, $T^{\pm z}$  split linearly with the magnetic field strength, as opposed to the  inter-valley coherent states, $T^{\pm x}$,  for which the opposite g-factors from the two valleys cancel.

Finally, one of the intriguing features of the computed spectra is the appearance of an orbitally symmetric \TP GS in a weakly gapped, weakly interacting bilayer. In the latter case, the short-range interaction (generally small compared to the Coulomb interaction and the resulting exchange splitting), determines the form of the  GS, and we find that inter-valley scattering at the lattice scale favors the formation of a Cooper-pair-like \TP state. This agrees with the earlier-discussed\cite{Lemonik2012} possibility of a singlet superconductor phasein BLG, which is a consequence of the same effectively attractive inter-valley interactions.

 In the limit of strong Coulomb repulsion (small $\epsilon$),   independently of the microscopic details or characteristics of the dot confinement, 
 the GS is  a four-fold degenerate multiplet where all spin and valley configurations ensure antisymmetry of the total wave function. For the excited states, depending on the dot  and the gap size, both orbitally antisymmetric, and orbitally symmetric states can occur, and in \fig\ref{fig:QD_Levelstructure} we manifest the  ordering of the low-energy states and  anticipated splittings in a weak perpendicular magnetic field for  typical hBN encapsulated BLG devices ($\epsilon\approx 4$\cite{Levinshtein2001, Rytova2018}) used in the recent experimental studies of BLG quantum dots\cite{Eich2018c, Eich2018a, Banszerus2019a}.

 \begin{acknowledgments}
We would like to thank A. Kurzmann, P. Rickhaus, M. Eich, K. Ensslin, T. Ihn, R. Kraft, R. Danneau, C. Stampfer, L. Banszerus, S. Slizovskiy, and B. Altshuler for discussions. We acknowledge funding from the European Graphene Flagship Project, the European Quantum Technology Project 2D-SIPC, the ERC Synergy Grant Hetero2D, EPSRC grants EP/S030719/1 and EP/N010345/1, and the Lloyd Register Foundation Nanotechnology Grant.
\end{acknowledgments}

\bibliography{QD}

\newpage

\setcounter{equation}{0}
\setcounter{figure}{0}
\renewcommand{\thefigure}{S\arabic{figure}}
\renewcommand{\theequation}{S\arabic{equation}}
 \renewcommand{\thesection}{S\arabic{section}}

\section*{Supplementary information for "Quartet states in two-electron quantum dots in bilayer graphene"}
  
  \section{Numerical Diagonalisation of the Single Particle Hamiltonian}

We can diagonalize the Hamiltonian,  $H_{\pm}^0$, in \eqn (5) numerically in  different bases of localised states. We choose the eigenstates of the two-dimensional  harmonic oscillator  (products of wave functions $\psi_n(x)=N_n e^{-\frac{1}{2}(\alpha x)^2}\mathcal{H}_n(\alpha x)$, where $N_n= \sqrt{\frac{\alpha}{\sqrt{\pi}2^n n!}}$ is the normalization constant and $\alpha$ is a scaling factor of unit length$^{-1}$; we choose $\alpha$ adapted to the potential $U(x)$ obtained from the fit of a parabolic potential to the bottom of $U$). The basis states are then given by 
\begin{equation}\psi_{\eta\mu ,1}
\nonumber\begin{pmatrix}
\psi_{\eta}(x)\psi_{\mu}(y)\\
0\\
0\\
0
\end{pmatrix},\;
\psi_{\eta\mu,2}
\begin{pmatrix}
0\\
\psi_{\eta}(x)\psi_{\mu}(y)\\
0\\
0
\end{pmatrix},
\end{equation}
\begin{equation}
\psi_{\eta\mu,3}
\begin{pmatrix}
0\\
0\\
\psi_{\eta}(x)\psi_{\mu}(y)\\
0
\end{pmatrix},
\psi_{ \eta\mu,4}
\begin{pmatrix}
0\\
0\\
0\\
\psi_{\eta}(x)\psi_{\mu}(y).
\end{pmatrix}.
 \label{eqn:BasisHarm}
\end{equation}

Alternatively, we can use the basis of eigenstates of 2D circular polar confinement in polar coordinates 
$\psi_{nm}(r,\varphi)=\mathcal{R}_{nm}(r)\;\mathcal{Y}_m(\varphi)$, where
\begin{align}
\nonumber \mathcal{R}_{nm}(r)&= N_{nm} (\alpha\,r)^{|m|}\,\mathcal{L}[n,|m|, \alpha^2 r^2]\, e^{-\frac{\alpha^2\,r^2}{2}},\\
\mathcal{Y}_m(\varphi)&= e^{im\varphi},
\label{eqn:BasisSpher}
\end{align}
where $N_{nm} =\alpha\sqrt{\frac{n!}{\pi(n+|m|!)}}$ ensures normalization and $\mathcal{L}$ denotes the associated Laguerre polynomials. \\

For every set of system parameters we construct the matrix corresponding to Hamiltonian $H^{0}_{\pm}$ in the basis given in \eqn\eqref{eqn:BasisHarm} or  \eqn\eqref{eqn:BasisSpher}  and obtain the energy spectrum by diagonalization. Convergence is reached when the energy levels do not change anymore upon including a higher number of basis states.

\section{Single Particle Spectra}

\begin{figure}[t!]
 \centering
\includegraphics[width=0.7\linewidth]{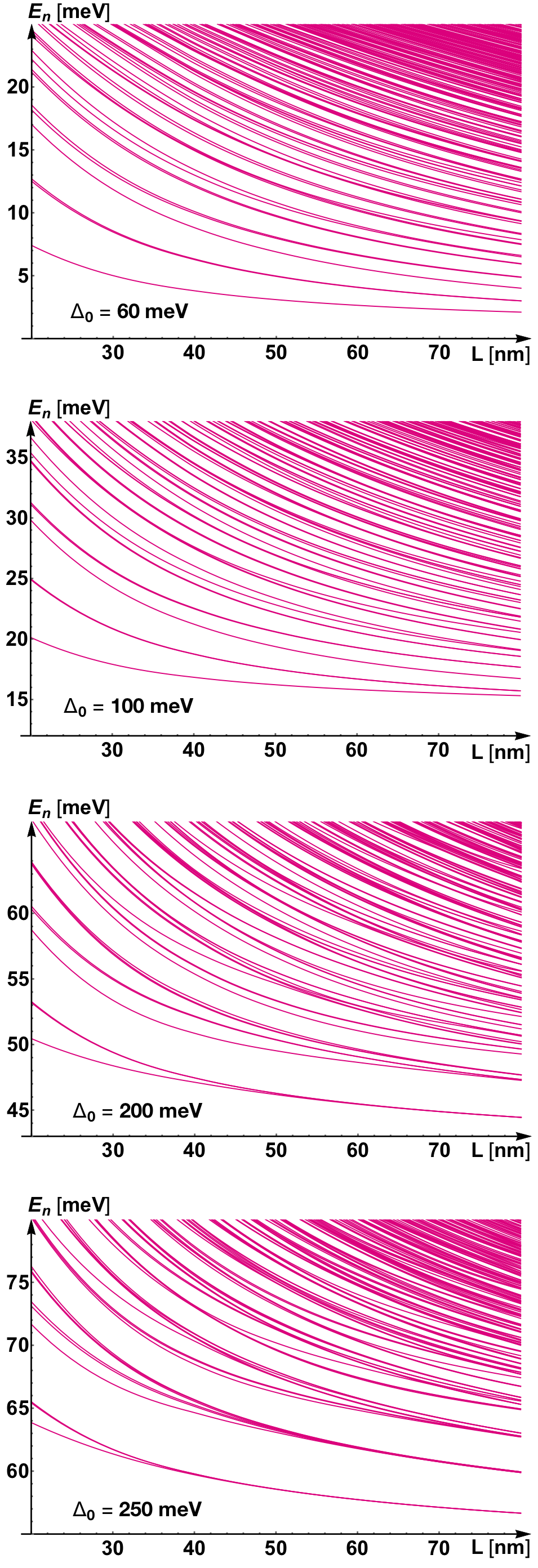}
\caption{ Single-particle QD spectra for different values of $\Delta_0$ as a function of $L$.}% Bottom panel: addition energies $\Delta E = E_{2P}-E_{SP}$ as a function of the dielectric $\epsilon$.}
\label{fig:SpectraDiffDwithL}
\end{figure}
\begin{figure}[t!]
\centering
\includegraphics[width=0.7\linewidth]{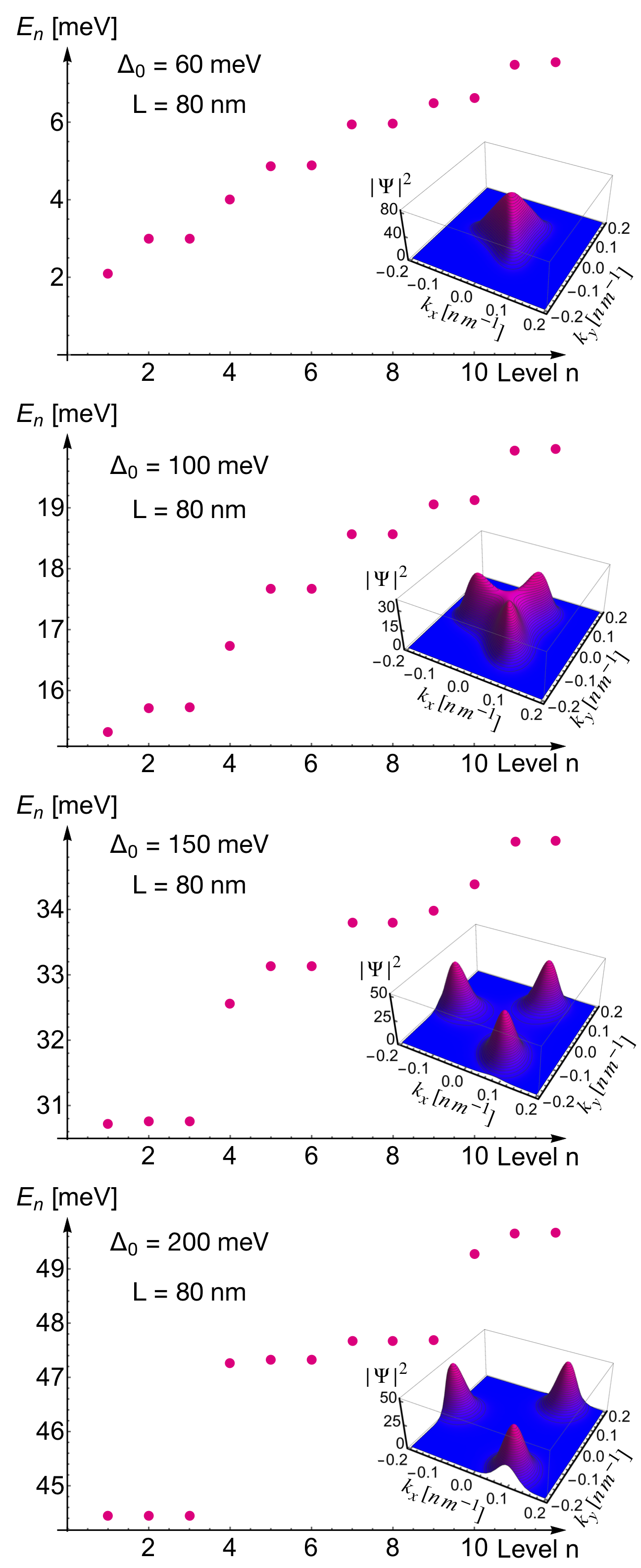}
\caption{ QD energy spectra for $L=80$ nm and different $\Delta_0$.  The insets show the probability distribution $|\Psi|^2$ in valley $K^-$ for the lowest level $n=1$. }% Bottom panel: addition energies $\Delta E = E_{2P}-E_{SP}$ as a function of the dielectric $\epsilon$.}
\label{fig:SpectraDiffD}
\end{figure}

\begin{figure}[t!]
 \centering
\includegraphics[width=0.75\linewidth]{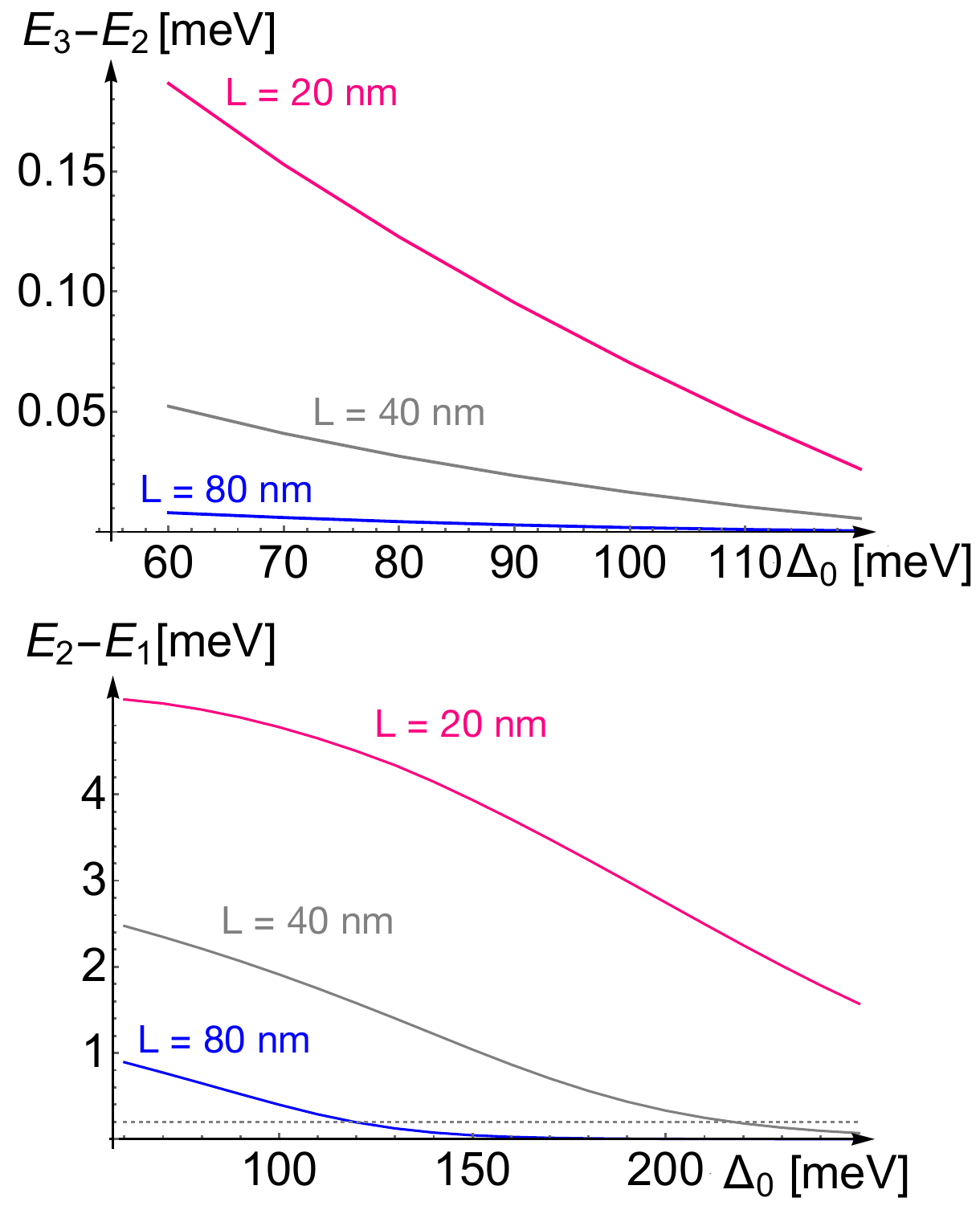}
\caption{ Properties of the spectra of symmetric QDs in gapped \BLG . Upper panel: distance between the third and second level, $E_3-E_2$, as a function of $\Delta_0$ for different $L$. Lower panel: distance between the second and first level, $E_2-E_1$, as a function of $\Delta_0$ for different $L$.}% Bottom panel: addition energies $\Delta E = E_{2P}-E_{SP}$ as a function of the dielectric $\epsilon$.}
\label{fig:SpectraStats}
\end{figure}

In \figs\ref{fig:SpectraDiffDwithL} and \ref{fig:SpectraDiffD} we show additional examples of \SP QD spectra. Figure \ref{fig:SpectraDiffDwithL}  illustrates the evolution of the \SP levels as a function of dot size, $L$, for different values of $\Delta_0$. Figure \ref{fig:SpectraDiffD} depicts in  detail the development from a singlet \SP GS (small $\Delta_0$), which is clearly separated from the higher energy states,  to a three-fold degenerate mini-valley GS (for large $\Delta_0$). In \fig \ref{fig:SpectraStats} we quantify the splittings of respective multiplets: The upper panel shows  how the splitting of the third and the second energy level, $E_3 - E_2$ develops as a function of $\Delta_0$ for different values of $L$. Since $\gamma_M$ (as discussed in the main text) is most substantial for small gaps, these two levels split the strongest for small $\Delta_0$ and small dotsTherefore, the splitting decreases with increasing $\Delta_0$ and increasing $L$. Further, we investigate the splitting of the first and the second energy level, $E_2 - E_1$, as a function of $\Delta_0$ for different values of $L$ (lower panel). 

\begin{figure}[t!]
\begin{minipage}{1.0\textwidth}
 \centering
\includegraphics[width=0.9\linewidth]{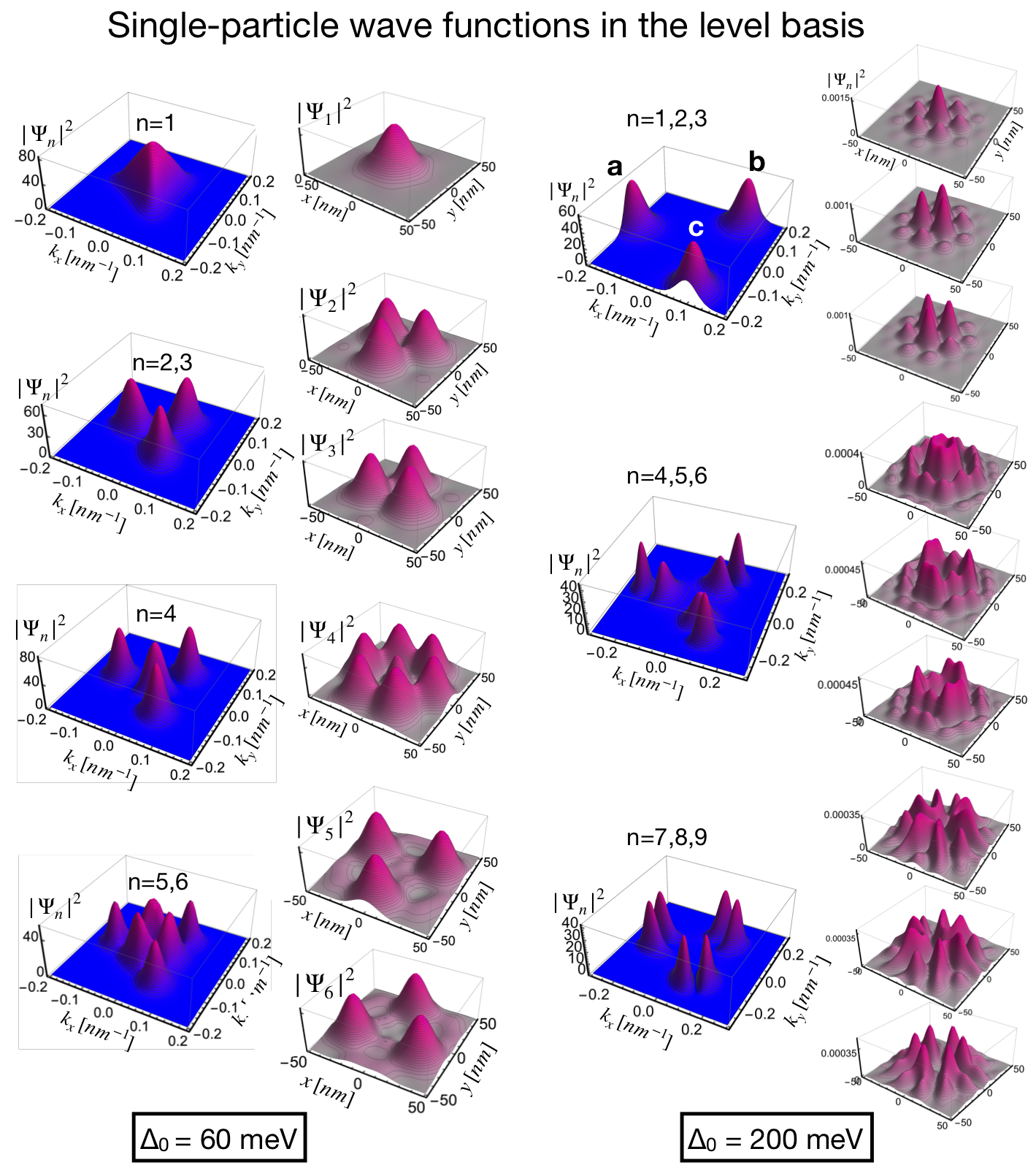}
\caption{ Momentum space  and real space probability distributions for the lowest six ($\Delta_0=60$ meV, left), or lowest nine ($\Delta_0=200$ meV, right) dot levels for a dot with L=80 nm, $\Delta_0=200$meV. These are the states used for the construction of the 2ES.}% Bottom panel: addition energies $\Delta E = E_{2P}-E_{SP}$ as a function of the dielectric $\epsilon$.}
\label{fig:NBasis}
\end{minipage}
\end{figure}

\begin{figure}[h!]
\begin{minipage}{1.0\textwidth}
\centering
\includegraphics[width=0.8\linewidth]{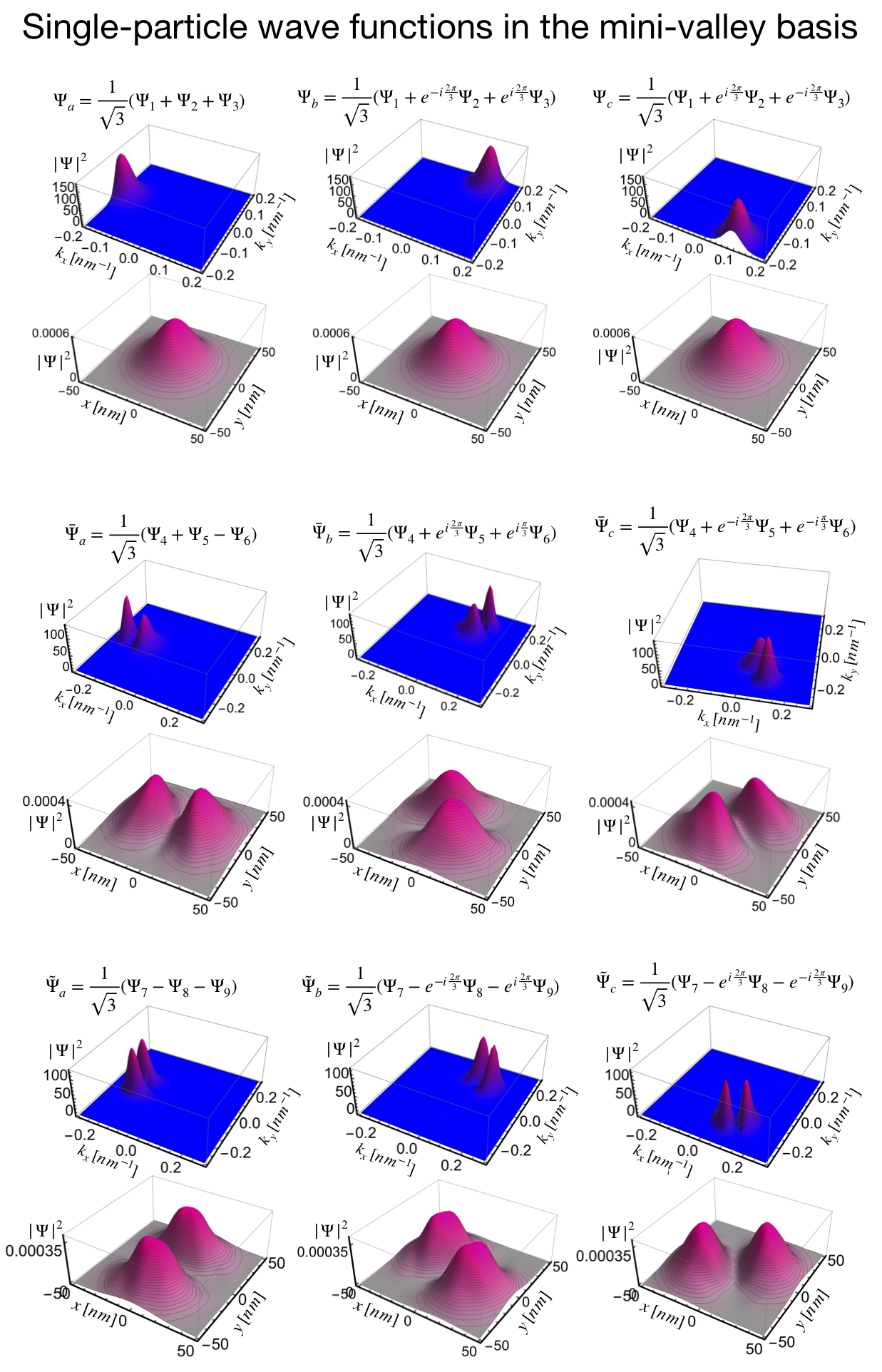}
\caption{ Transformation  from the level basis, $\SPstate_n$, to the basis of mini-valley states, labelled by $a,b,c$ in a strongly gapped BLG QD. We illustrate the probability distributions in momentum space (top rows) and in real space (bottom space) for a dot with L=80 nm, $\Delta_0=200$meV. }
\label{fig:MVBasis}
\end{minipage}
\end{figure}

The regime in which the levels of the spectra are each threefold degenerate (corresponding to the three mini-valleys around each valley) is realized for large dots and large gaps. In this case, the minivalleys around each of BLG's valleys is sufficiently developed and seperated in momentum space to approximately represent a good quantum number. We can hence express the \SP dot levels in a minivalley basis, where $a,b,c$ label the three minivalleys and  $\SPstate_{a,b,c}$  denotes the minivalley states of the lowest \SP triplet, while $\bar{\SPstate}_{a,b,c}$ and  $\tilde{\SPstate}_{a,b,c}$ refer to those of the second and third triplet respectively. The change of basis is illustrated in \fig\ref{fig:MVBasis} where we show the probability distributions of all the mini-valley states both in momenutm space and in real space.  For comparison, in \fig\ref{fig:NBasis}, we show equally the momentum space and real space distributions of the corresponding nine  lowest dot states, $n=1,\dots,9$ in the level basis $\SPstate_n$.

\clearpage
\clearpage

\begin{figure}[t!]
 \centering
\includegraphics[width=0.7\linewidth]{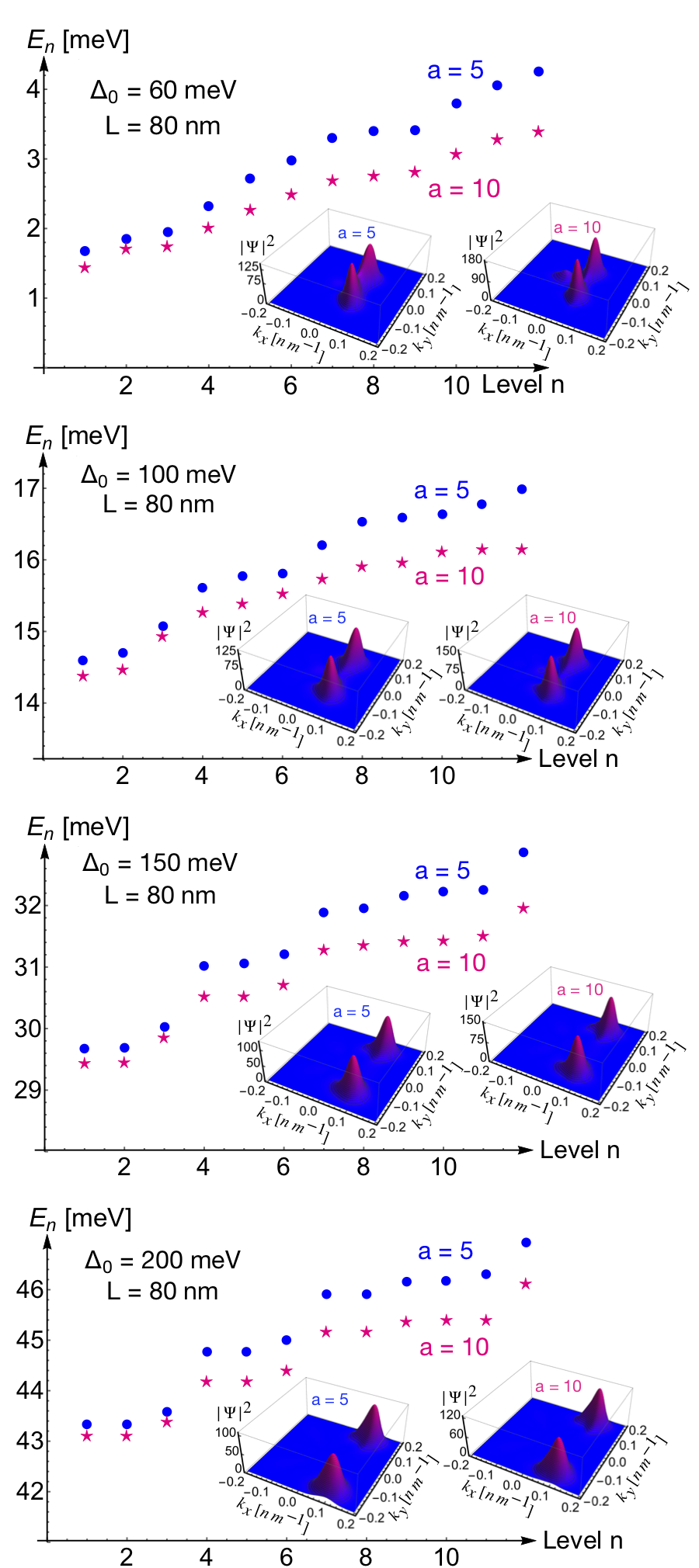}
\caption{ BLG QD spectra for elliptical QDs with $a=5$ (blue circles) or $a=10$ (magenta stars) for $L=80$ nm and different $\Delta_0$. The insets show the probability distribution $|\Psi|^2$ in valley $K^-$ for the wave functions of the lowest level $n=1$, respectively.}% Bottom panel: addition energies $\Delta E = E_{2P}-E_{SP}$ as a function of the dielectric $\epsilon$.}
\label{fig:SpectraEllDiffD}
\end{figure}

Furthermore, we mention the generalization of the model discussed in the main text, when the dot confinement is elliptical with confining potential and  gap  of the form
\begin{equation}
U(x,y)=\frac{U_0}{\cosh{\frac{\sqrt{x^2+ \frac{y^2}{a}}}{L}}}, \;\Delta(x,y)=\Delta_0-\frac{\beta\Delta_0}{\cosh{\frac{\sqrt{x^2+ \frac{y^2}{a}}}{L}}}.
\label{eqn:Pot}
\end{equation}
The types of confinement described range from the familiar circular symmetric confinement (for $a=1$) or a potential of elliptical shape  (elongated along the $y$-direction for $a>1$). Examples for \SP spectra of elliptical QDs are depicted in \fig\ref{fig:SpectraEllDiffD}. 

\section{Influence of trigonal warping on   dispersion and   Berry curvature}

\begin{figure}[t!]
 \centering
\includegraphics[width=0.7\linewidth]{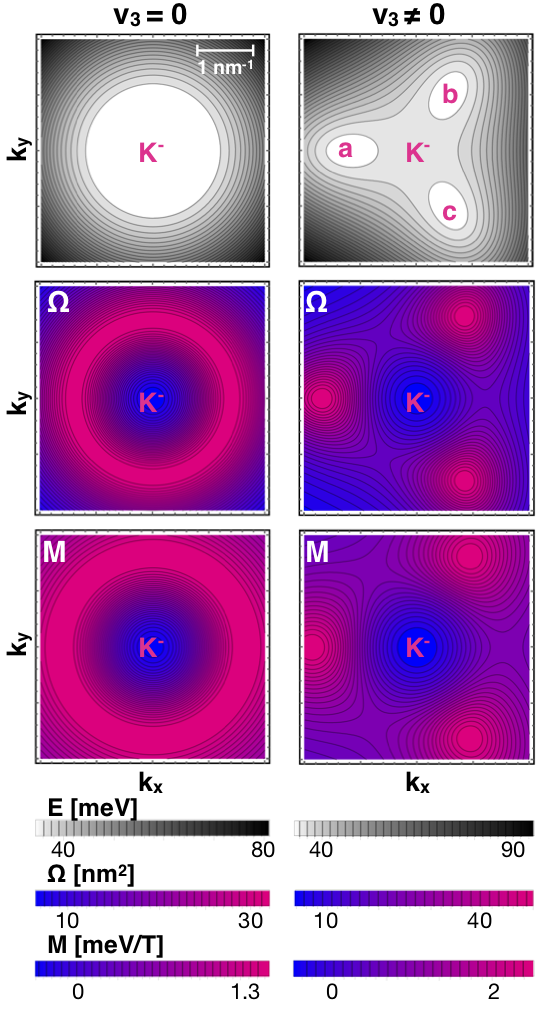}
\caption{ Electronic structure of homogeneous gapped \BLG in the absence of confinement with or without trigonal warping effects  (left column, $v_3\equiv0$, or right column, $v_3= 0.12v$). Top panels: electron dispersion of the lowest conduction band. Centre and bottom panels: Berry curvature, $ \Omega(\mathbf{k})$, and orbital magnetic moment, $\mathbf{M}(\mathbf{k})$, computed numerically  for the lowest conduction band and $\Delta=70$ meV. In the absence of $v_3$ the electronic structure has rotational symmetry around the $K$ point, while the trigonally warped structure exhibits three minivalleys in the dispersion (labelled a,b, and c), or three peaks in $\Omega$ and $M$, respectively, around each $K$ point. In the $K^+$ valley, the structures in momentum space are rotated by $\pi$ and both,  $\Omega$ and $M$, carry opposite sign in the opposite valley.}% Bottom panel: addition energies $\Delta E = E_{2P}-E_{SP}$ as a function of the dielectric $\epsilon$.}
\label{fig:QD_BLGEOmM}
\end{figure}

From the Bloch functions of BLG, we obtain the corresponding Berry curvature, $ \Omega(\mathbf{k})$ , and orbital magnetic moment, $\mathbf{M}(\mathbf{k})=M(\mathbf{k})\mathbf{e}_z$, as \cite{Xiao2010, Chang1996a}
\begin{align}
\nonumber{\Omega}&=i \LD\mathbf{\nabla}_{\mathbf{k}}\Phi(\mathbf{k})|\times|\mathbf{\nabla}_{\mathbf{k}}\Phi(\mathbf{k})\RD\cdot\mathbf{e}_z ,\\
{M}&=-i\frac{e}{2\hbar}\LD\mathbf{\nabla}_{\mathbf{k}}\Phi(\mathbf{k})|\times [\epsilon(\mathbf{k})-H(\mathbf{k})] |\mathbf{\nabla}_{\mathbf{k}}\Phi(\mathbf{k})\RD\cdot\mathbf{e}_z,
\label{eqn:Berry}
\end{align}
where, $\mathbf{\nabla}_{\mathbf{k}}=(\partial_{k_x},\partial_{k_y})$, $"\times"$ is the cross product, and $\epsilon(\mathbf{k})$ is the band energy.  Both quantities, $\Omega$ and $M$, inherit the threefold rotational symmetry of the trigonally warped bands:   $M$ and $\Omega$ develop non-zero peaks around the positions of the minivalleys in momentum space, where both,  $\Omega$ and $M$, carry opposite sign in the opposite valley. Note that the Berry curvature and the orbital magnetic moment are closely related quantities. In fact, within the two-band model of \BLG  \cite{McCann2006, McCann2007} with $v_3\rightarrow0$ \cite{Park2017, Fuchs2010}, for which $\epsilon(\mathbf{k})=\sqrt{(\frac{\hbar^2 k^2}{2m})^2+(\frac{\Delta}{2})^2}$ with $m\approx\frac{\gamma_1}{2v^2}\approx0.032m_e$ being the effective mass of electrons in BLG, they are directly proportional: $\Omega\approx-\xi\frac{\hbar^2}{2m}\frac{\hbar^2 k^2}{2m}\frac{\Delta}{\epsilon(\mathbf{k})^3}$ and $M\approx  -\epsilon(\mathbf{k}) \frac{e}{\hbar}\Omega=\xi\frac{e\hbar}{2m}\frac{\hbar^2 k^2}{2m}\frac{\Delta}{\epsilon(\mathbf{k})^2}$ , where $\xi=\pm1$ labels the two different valleys.

%\begin{figure}[t!]
 %\centering
 %\begin{minipage}{1.0\textwidth}
%\includegraphics[width=1.0\linewidth]{SpectrumNoTrigowWFwM}
%\caption{ Energy Levels and wave functions of a circular isotropic \BLG QD in the absence of trigonal warping for $\Delta_0=100$ meV and $L=80$ nm. For each level, we show the corresponding $m$ quantum number and the probability distribution in momentum space for the wave function on the second lattice atom,$|\Psi|^2$ in valley $ K^-$. This plot is for the  $K^{-}$ valley, in the  $K^{+}$ valley the picture is identical with the sign of $m$ in each level is reversed and the wave functions are rotated by $\pi$.}% Bottom panel: addition energies $\Delta E = E_{2P}-E_{SP}$ as a function of the dielectric $\epsilon$.}
%\label{fig:SpectrumNoTrigowWFwM}
%\end{figure*}
%\begin{figure*}[t!]
 %\centering
 %\end{minpage}
 %\begin{minpage}{1.0\textwidth}
%\includegraphics[width=1.0\linewidth]{SpectrumTrigowWFwM}
%\caption{ Same as \fig \ref{fig:SpectrumNoTrigowWFwM} for the case when trigonal warping is taken into account. }% Bottom panel: addition energies $\Delta E = E_{2P}-E_{SP}$ as a function of the dielectric $\epsilon$.}
%\label{fig:SpectrumTrigowWFwM}
%\end{minipage}
%\end{figure}

\begin{figure}[t!]
 \begin{minipage}{1.0\textwidth}
\includegraphics[width=1.0\linewidth]{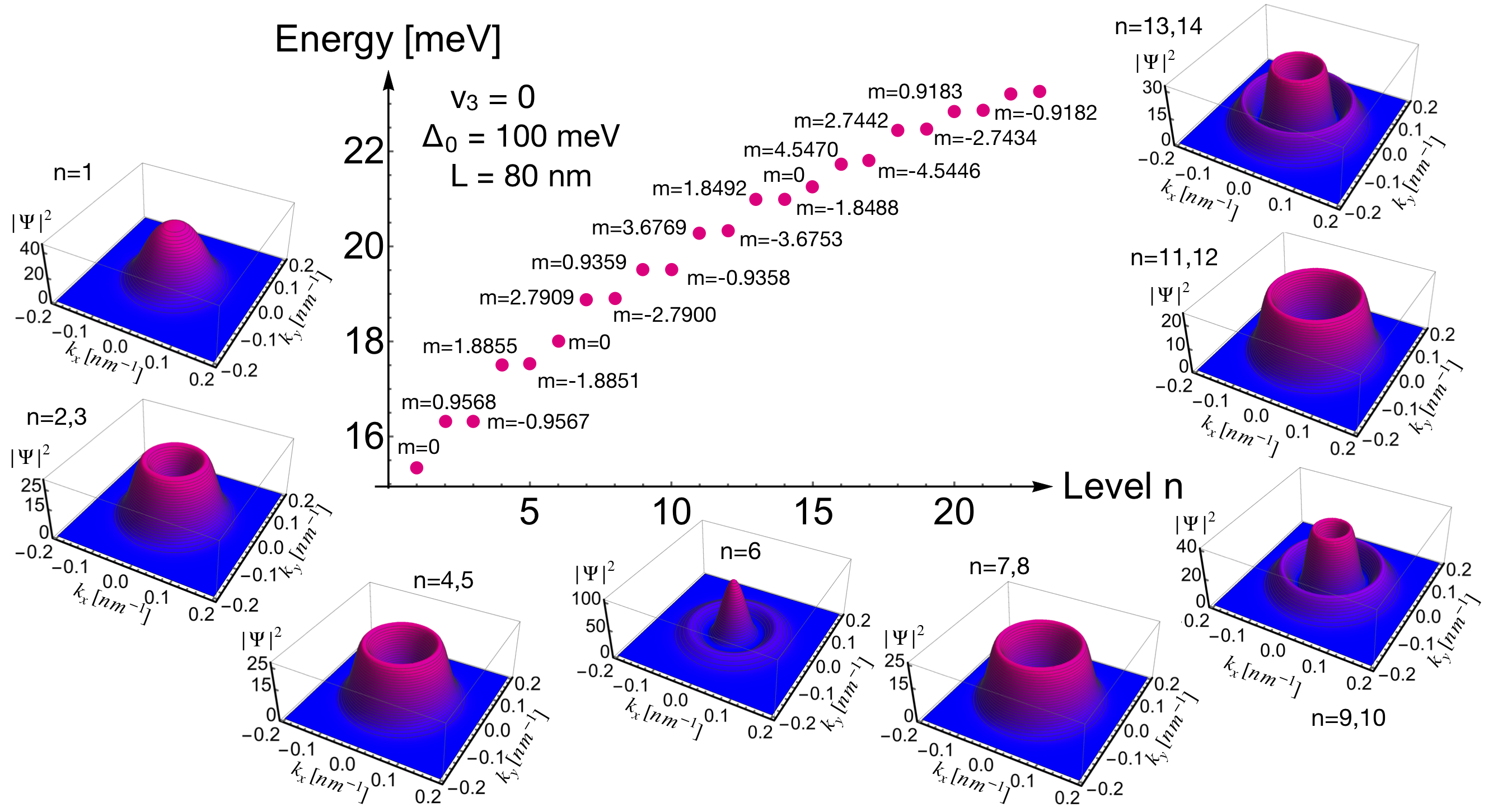}
\caption{ Energy Levels and wave functions of a circular isotropic \BLG QD in the when neglecting trigonal warping for $\Delta_0=100$ meV and $L=80$ nm. For each level, we show the corresponding $m$ quantum number and the probability distribution in momentum space for the wave function on the second lattice atom,$|\Psi|^2$ in valley $ K^-$. This plot is for the  $K^{-}$ valley, in the  $K^{+}$ valley the picture is identical with the sign of $m$ in each level is reversed and the wave functions are rotated by $\pi$.}% Bottom panel: addition energies $\Delta E = E_{2P}-E_{SP}$ as a function of the dielectric $\epsilon$.}
\label{fig:SpectrumNoTrigowWFwM}
 \end{minipage}
 \begin{minipage}{1.0\textwidth}
\includegraphics[width=1.0\linewidth]{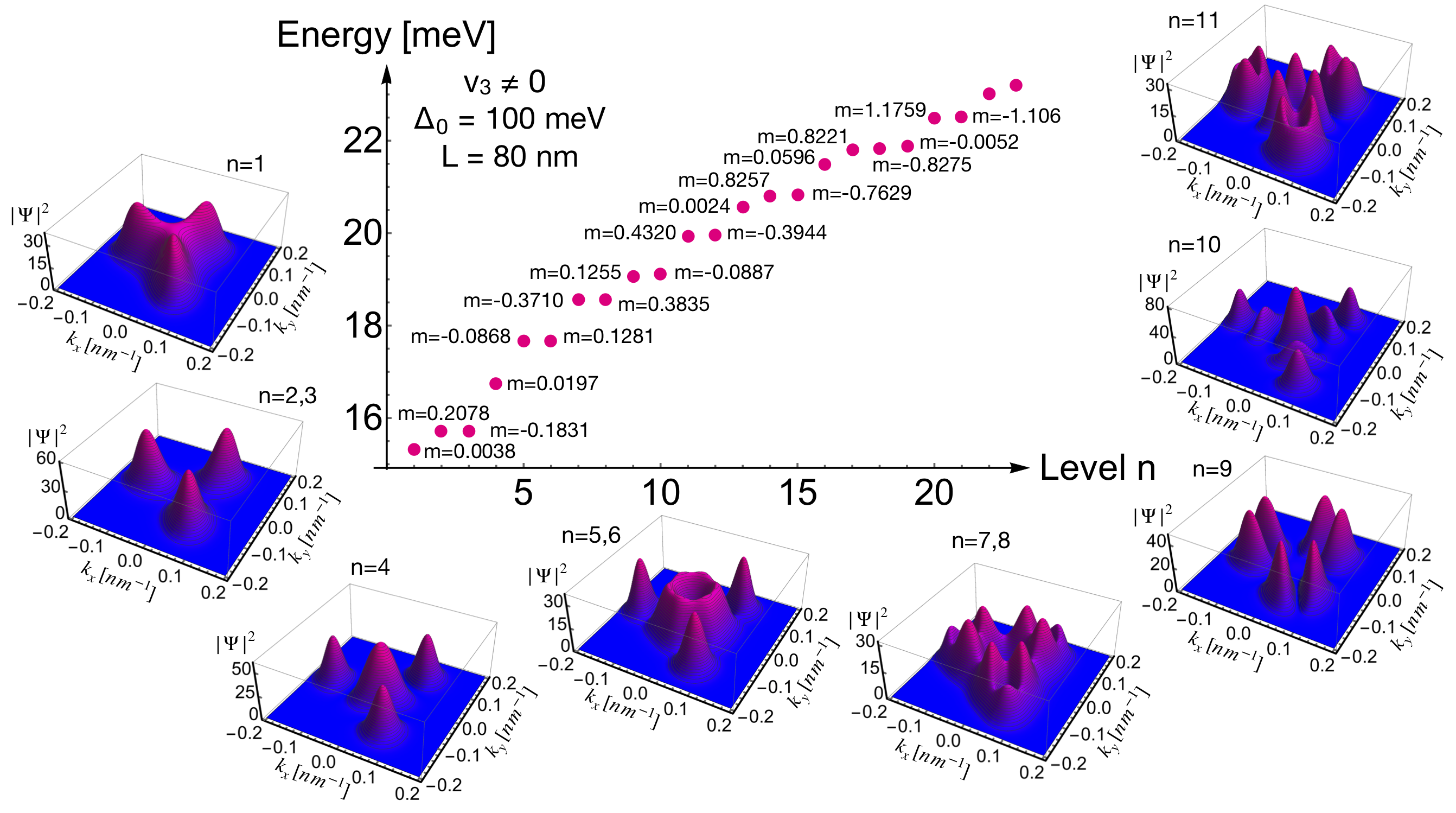}
\caption{ Same as \fig \ref{fig:SpectrumNoTrigowWFwM} for the case when trigonal warping is taken into account. }% Bottom panel: addition energies $\Delta E = E_{2P}-E_{SP}$ as a function of the dielectric $\epsilon$.}
\label{fig:SpectrumTrigowWFwM}
\end{minipage}
\end{figure}

 We comment on how trigonal warping, induced by $v_3$, breaks continuous rotational symmetry and reduces the symmetry of the dispersion (and, consequently, of the Berry curvature and the orbital magnetic moment) to $C_3$. We compare the quantities computed without trigonal warping ($v_3\equiv0$) and with trigonal warping ($v_3\neq0$) in \fig\ref{fig:QD_BLGEOmM}. To illustrate the effect of the symmetry breaking on the BLG QD spectra and states we show both examples, in the absence and in the presence of trigonal warping, in \figs\ref{fig:SpectrumNoTrigowWFwM} and \ref{fig:SpectrumTrigowWFwM}, respectively. 

In the absence of any symmetry breaking, electrons subject to harmonic circularly isotropic confinement without a magnetic field exhibit quantization into discrete Fock-Darwin energy levels\cite{Fock1928, Darwin1931},
\begin{equation}
E_{l,m}=(2l+|m|+1)\hbar\omega_0,
\label{eqn:FockDarwin}
\end{equation}
with a characteristic frequency $\omega_0$ and characterised by  two quantum numbers, $l\in\mathbb{N}^{0}$ (radial quantum number), and $m\in\mathbb{Z}$ (angular momentum quantum number).

To demonstrate the effect of the additional orbital angular momentum due to the non-trivial Berry curvature of the states in gapped BLG, it is instructive to consider in some detail the system in the absence of trigonal warping, \textit{i.e.}, for $v_3\equiv0$. In this case, when both the confinement and the dispersion exhibit rotational symmetry (see top left panel of \fig\ref{fig:QD_BLGEOmM}), the additional angular momentum due to the Berry curvature represents the only perturbation to the usual Fock-Darwin levels as given in \eqn\ref{eqn:FockDarwin}.

In \fig\ref{fig:SpectrumNoTrigowWFwM} we show an example of a spectrum for circular confinement in gapped \BLG if trigonal warping is neglected for $\Delta_0=100$ meV and $L=80$ nm. We show the probability density for the wave functions, $|\Psi |^2$ in valley $ K^-$  and indicate for each level the value of the $m$ quantum number of this state. This spectrum has been calculated for the $K^{-}$ valley,  an equivalent picture is obtained for the $K^{+}$ valley with signs of each $m$ reversed. We hence find degeneracy between states with $(K=+1, m)$ and $(K=-1, -m)$. The $m$ quantum number is still close to the integer values prescribed by the Fock-Darwin states  with small perturbations due to additional effects by the additional orbital angular momentum influencing the angular momentum of the states. Furthermore, we see from  \fig\ref{fig:SpectrumNoTrigowWFwM} that, while the spectrum and the wave functions retain some of the properties of the Fock-Darwin levels in zero magnetic field, all degeneracies between levels with different $m$ in one valley are lifted. Previously exactly degenerate levels with quantum numbers $\pm |m|$ are split apart slightly due to coupling between the orbital magnetic moment, $\mathbf{\Omega}$, and the angular momentum $\mathbf{L}_z$ of the form $\mathbf{\Omega}\cdot\mathbf{L}_z$. The splitting of these pairs increases with increasing $|m|$. In cases where the previously degenerate state carries a different quantum number $m$ (\textit{i.e.~} the triplet \{$l=1$, $m=0$\}, \{$l=0$, $m=\pm2$\} ) it is split apart even more strongly from its previously degenerate partners. Due to the different distribution of the wavefunctions in momentum space, states with different $|m|$ pick up different amounts of Berry curvature and therefore carry different orbital magnetic moment. These splittings can be found to be directly proportional to the difference in Berry curvature of states with different $|m|$.

\clearpage
\clearpage

If $v_3\neq0$ is taken into account, the \BLG dispersion acquires the trigonally warped dispersion shown in the top right panel of \fig\ref{fig:QD_BLGEOmM}. Therefore, in this case, rotational symmetry is broken even in the presence of  rotationally symmetric confinement. In \fig\ref{fig:SpectrumTrigowWFwM} we demonstrate the effect of trigonal warping on the energy levels, the characterising quantum numbers $m$, and the corresponding wave functions. Degeneracies are lifted, now due to both, Berry curvature effects and breaking of rotational symmetry. The wave functions exhibit a complex pattern with modes emerging in each mini-valley, separately. As a consequence, the $m$-values do no longer resemble the integer values of the Fock-Darwin levels.

\section{Coupling to an Electric Field and Optical Transition Matrix Elements}

\begin{figure}[t!]
 \centering
\includegraphics[width=1\linewidth]{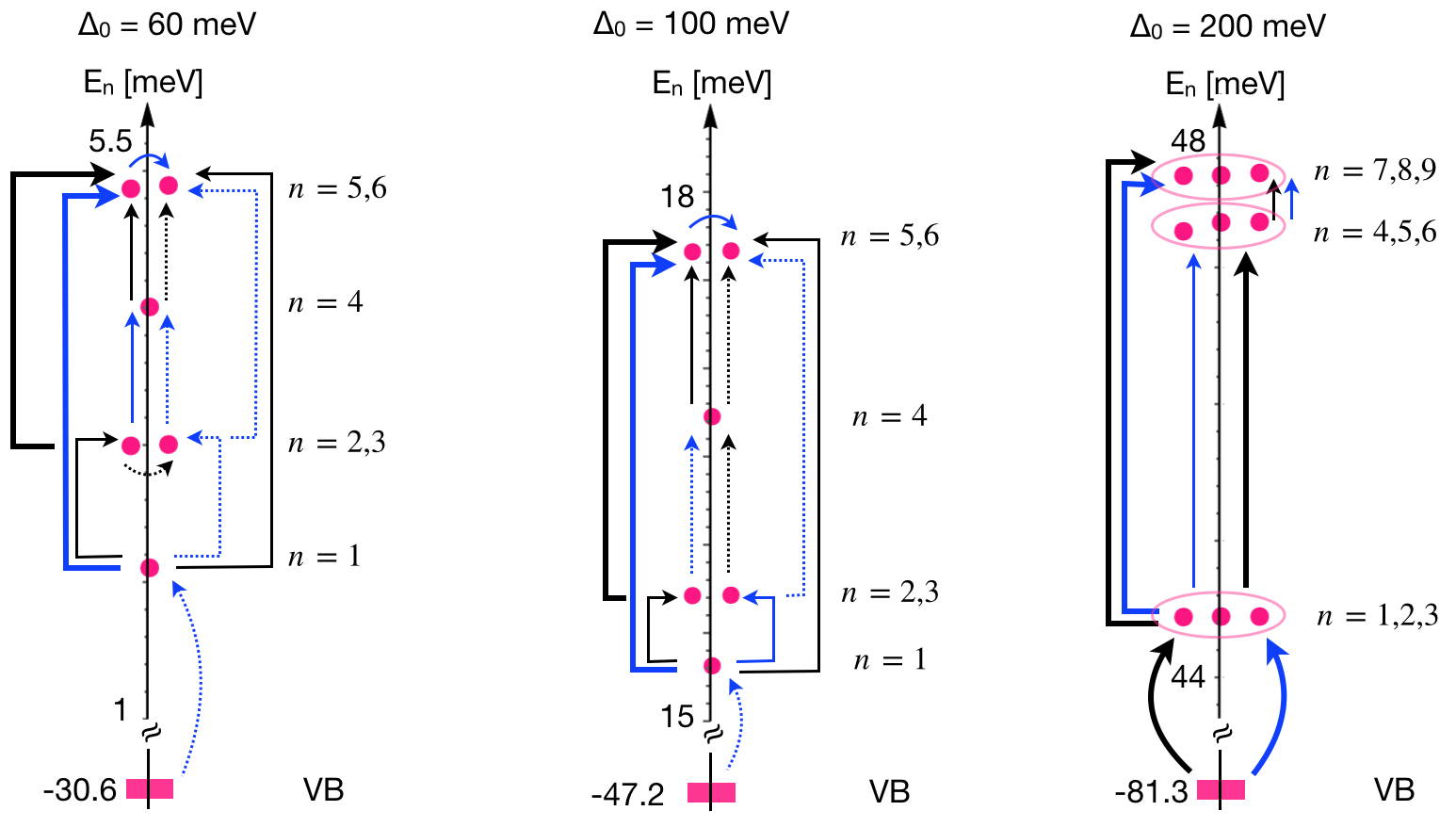}
\caption{Channels for optical absorption of  right ($\boldsymbol{\ell}_{\circlearrowright}$, blue arrows) and left-handed ($\boldsymbol{\ell}_{\circlearrowleft}$, black arrows) circularly polarized light in the $K^+$ valley. Different arrows indicate different strengths of the transitions, quantified by the transition matrix elements $t^+_{n^{\prime},n} $ : thick line: $t^+_{n^{\prime},n}>1$, thin line: $0.1<t^+_{n^{\prime},n}<1$, dotted line:    $0<t^+_{n^{\prime},n}<0.1$ (in units of $\frac{meV}{E_{\omega}[mV/nm]/\hbar\omega[meV]} $)  .}% Bottom panel: addition energies $\Delta E = E_{2P}-E_{SP}$ as a function of the dielectric $\epsilon$.}
\label{fig:OpticsAll3}
\end{figure}

The lowest order term describing coupling between the electrons in \BLG and the electromagnetic field (introduced via Peierl's substitution)  for small momenta  $\mathbf{p}$ is given by \cite{Mucha-Kruczynski2009, Mucha-Kruczynski2012}
\begin{equation}
H_{e-em}^{\pm}=\mathbf{A} \cdot\mathbf{j}^{\pm}, 
\end{equation}
with the current operator
\begin{equation}
\mathbf{j}^{\pm}=-e\frac{\partial H^{0}_{\pm}(\mathbf{p})}{\partial \mathbf{p}}.
\end{equation}
The incoming light is described by an electric field (neglecting the momentum of the photon compared to that of the electronic states), $\mathbf{E}_{\omega}=\mathbf{E}_{\omega}e^{-i\omega t}$. Using Maxwell's equations, $\mathbf{\mathbf{E}}=-\frac{\partial\mathbf{A}}{\partial t}$ (for the transverse field modes which couple to the electronic current), we can write
\begin{equation}
\mathbf{A} = \frac{1}{i\omega}\mathbf{E}_{\omega}e^{-i\omega t}.
\end{equation}
We use the numerical wave functions of the electrons in the \BLG QD to calculate the matrix elements that determine the rules for transitions between the QD levels. Transitions are induced by absorption of right ($\circlearrowright$) and left-handed ($\circlearrowleft$) circularly polarized light $\mathbf{E}_{\omega}={E}_{\omega}\boldsymbol{\ell}_{\circlearrowright/\circlearrowleft}$, where $\boldsymbol{\ell}_{\circlearrowright/\circlearrowleft}=\frac{1}{\sqrt{2}}(\mathbf{e}_x\mp \mathbf{e}_y)$, and $\mathbf{e}_{x/y}$ denote unit vectors. We choose the examples $\Delta_0=60$,  $\Delta_0=100$, and  $\Delta_0=200$ meV to illustrate the optical absorption spectra in \fig\ref{fig:OpticsAll3}. The values for the transition matrix elements in the $K^+$ valley,
\begin{equation}
t^+_{n^{\prime},n} = \LD n^{\prime}| H^{+}_{e-em,\ell_{\circlearrowright/\circlearrowleft}}| n\RD ,
\end{equation}
 for  different combinations of initial states $n$ and final states $n^{\prime}$  are summarized in tables \ref{tab:SR_TrigoD60}, \ref{tab:SR_TrigoD100}, \ref{tab:SR_TrigoD200}, respectively, in units of $\frac{meV}{E_{\omega}[mV/nm]/\hbar\omega[meV]} $.

\clearpage
 \begin{minipage}{1.0\textwidth}
%\begin{widetext}
\begin{equation}
\resizebox{1\hsize}{!}{$
\begin{array}{|c|c|c|c|c|c|}\hline   \Delta_0=60\; meV& & & & &\\
  \hline
  \LD1|H_{int,\ell_{\circlearrowright}}|1\RD = 0 &  \LD2|H_{int,\ell_{\circlearrowright}}|1\RD = 0 &  \LD3|H_{int,\ell_{\circlearrowright}}|1\RD =0.09 & \LD4|H_{int,\ell_{\circlearrowright}}|1\RD =0 & \LD5|H_{int,\ell_{\circlearrowright}}|1\RD =-0.82 & \LD6|H_{int,\ell_{\circlearrowright}}|1\RD =0  \\
    \LD1| H_{int,\ell_{\circlearrowleft}}|1\RD = 0 &  \LD2| H_{int,\ell_{\circlearrowleft}}|1\RD = -0.36&  \LD3| H_{int,\ell_{\circlearrowleft}}|1\RD =0 & \LD4|\ H_{int,\ell_{\circlearrowleft}}|1\RD =0 & \LD5| H_{int,\ell_{\circlearrowleft}}|1\RD =0 & \LD6| H_{int,\ell_{\circlearrowleft}}|1\RD =0.15\\
    \hline
      \LD1|H_{int,\ell_{\circlearrowright}}|2\RD =  -0.36 &  \LD2|H_{int,\ell_{\circlearrowright}}|2\RD =0   &  \LD3|H_{int,\ell_{\circlearrowright}}|2\RD =0  & \LD4|H_{int,\ell_{\circlearrowright}}|2\RD =0 & \LD5|H_{int,\ell_{\circlearrowright}}|2\RD = 0 & \LD6|H_{int,\ell_{\circlearrowright}}|2\RD = 0 \\
    \LD1| H_{int,\ell_{\circlearrowleft}}|2\RD = 0  &  \LD2| H_{int,\ell_{\circlearrowleft}}|2\RD = 0  &  \LD3| H_{int,\ell_{\circlearrowleft}}|2\RD = 0.29  & \LD4| H_{int,\ell_{\circlearrowleft}}|2\RD = 0.35 & \LD5| H_{int,\ell_{\circlearrowleft}}|2\RD = -1.32 & \LD6| H_{int,\ell_{\circlearrowleft}}|2\RD = 0\\
    \hline
     \LD1|H_{int,\ell_{\circlearrowright}}|3\RD =0     &  \LD2|H_{int,\ell_{\circlearrowright}}|3\RD =   0.29   &  \LD3|H_{int,\ell_{\circlearrowright}}|3\RD =   0& \LD4|H_{int,\ell_{\circlearrowright}}|3\RD =  0.04 & \LD5|H_{int,\ell_{\circlearrowright}}|3\RD =0  & \LD6|H_{int,\ell_{\circlearrowright}}|3\RD =-0.03  \\
    \LD1| H_{int,\ell_{\circlearrowleft}}|3\RD = 0.09 &  \LD2| H_{int,\ell_{\circlearrowleft}}|3\RD = 0   &  \LD3| H_{int,\ell_{\circlearrowleft}}|3\RD =  0  & \LD4| H_{int,\ell_{\circlearrowleft}}|3\RD =  0 & \LD5| H_{int,\ell_{\circlearrowleft}}|3\RD = 0 & \LD6| H_{int,\ell_{\circlearrowleft}}|3\RD = 0 \\
    \hline
     \LD1|H_{int,\ell_{\circlearrowright}}|4\RD =  0 &  \LD2|H_{int,\ell_{\circlearrowright}}|4\RD =0.35&  \LD3|H_{int,\ell_{\circlearrowright}}|4\RD = 0 & \LD4|H_{int,\ell_{\circlearrowright}}|4\RD =  0& \LD5|H_{int,\ell_{\circlearrowright}}|4\RD =0.47 &  \LD6|H_{int,\ell_{\circlearrowright}}|4\RD = 0\\
    \LD1| H_{int,\ell_{\circlearrowleft}}|4\RD = 0  &  \LD2| H_{int,\ell_{\circlearrowleft}}|4\RD =0    &  \LD3| H_{int,\ell_{\circlearrowleft}}|4\RD = 0.04 & \LD4| H_{int,\ell_{\circlearrowleft}}|4\RD =0  & \LD5| H_{int,\ell_{\circlearrowleft}}|4\RD = 0  & \LD6| H_{int,\ell_{\circlearrowleft}}|4\RD = -0.07 \\
    \hline
     \LD1|H_{int,\ell_{\circlearrowright}}|5\RD =  0 & \LD2 |H_{int,\ell_{\circlearrowright}}|5\RD =  -1.32 &  \LD3 |H_{int,\ell_{\circlearrowright}}|5\RD =  0 &  \LD4 |H_{int,\ell_{\circlearrowright}}|5\RD =  0 &  \LD5 |H_{int,\ell_{\circlearrowright}}|5\RD =  0  &  \LD6|H_{int,\ell_{\circlearrowright}}|5\RD =  0.31 \\  
      \LD1|H_{int,\ell_{\circlearrowleft}}|5\RD =  -0.82  &    \LD2|H_{int,\ell_{\circlearrowleft}}|5\RD = 0 &   \LD3|H_{int,\ell_{\circlearrowleft}}|5\RD = 0 &   \LD4|H_{int,\ell_{\circlearrowleft}}|5\RD = 0.47 &   \LD5|H_{int,\ell_{\circlearrowleft}}|5\RD = 0  &   \LD6|H_{int,\ell_{\circlearrowleft}}|5\RD = 0 \\  
      \hline 
\end{array}$}
\label{tab:SR_TrigoD60}
\end{equation}
\end{minipage}
%\end{widetext}
%\begin{widetext}
 \begin{minipage}{1.0\textwidth}
\begin{equation}
\resizebox{1\hsize}{!}{$
\begin{array}{|c|c|c|c|c|c|}\hline  \Delta_0=100\; meV& & & & &\\
  \hline
  \LD1|H_{int,\ell_{\circlearrowright}}|1\RD = 0 &  \LD2|H_{int,\ell_{\circlearrowright}}|1\RD = 0 &  \LD3|H_{int,\ell_{\circlearrowright}}|1\RD =0.40 & \LD4|H_{int,\ell_{\circlearrowright}}|1\RD =0 & \LD5|H_{int,\ell_{\circlearrowright}}|1\RD =1.22 & \LD6|H_{int,\ell_{\circlearrowright}}|1\RD =0\\
    \LD1| H_{int,\ell_{\circlearrowleft}}|1\RD = 0 &  \LD2| H_{int,\ell_{\circlearrowleft}}|1\RD =  0.12  &  \LD3| H_{int,\ell_{\circlearrowleft}}|1\RD =0 & \LD4|\ H_{int,\ell_{\circlearrowleft}}|1\RD =0 & \LD5| H_{int,\ell_{\circlearrowleft}}|1\RD =0 & \LD6|H_{int,\ell_{\circlearrowleft}}|1\RD =0.42 \\
    \hline
      \LD1|H_{int,\ell_{\circlearrowright}}|2\RD =  0.12  &  \LD2|H_{int,\ell_{\circlearrowright}}|2\RD =0   &  \LD3|H_{int,\ell_{\circlearrowright}}|2\RD =0  & \LD4|H_{int,\ell_{\circlearrowright}}|2\RD = 0.02 & \LD5|H_{int,\ell_{\circlearrowright}}|2\RD = 0  & \LD6|H_{int,\ell_{\circlearrowright}}|2\RD = 0\\
    \LD1| H_{int,\ell_{\circlearrowleft}}|2\RD = 0  &  \LD2| H_{int,\ell_{\circlearrowleft}}|2\RD = 0  &  \LD3| H_{int,\ell_{\circlearrowleft}}|2\RD = 0.59  & \LD4| H_{int,\ell_{\circlearrowleft}}|2\RD = 0 & \LD5| H_{int,\ell_{\circlearrowleft}}|2\RD = 1.85 & \LD6| H_{int,\ell_{\circlearrowleft}}|2\RD = 0\\
    \hline
     \LD1|H_{int,\ell_{\circlearrowright}}|3\RD =0     &  \LD2|H_{int,\ell_{\circlearrowright}}|3\RD =   0.59   &  \LD3|H_{int,\ell_{\circlearrowright}}|3\RD =   0& \LD4|H_{int,\ell_{\circlearrowright}}|3\RD =  0 & \LD5|H_{int,\ell_{\circlearrowright}}|3\RD =0   & \LD6|H_{int,\ell_{\circlearrowright}}|3\RD =-0.07  \\
    \LD1| H_{int,\ell_{\circlearrowleft}}|3\RD =  0.40  &  \LD2| H_{int,\ell_{\circlearrowleft}}|3\RD = 0   &  \LD3| H_{int,\ell_{\circlearrowleft}}|3\RD =  0  & \LD4| H_{int,\ell_{\circlearrowleft}}|3\RD =  0.07 & \LD5| H_{int,\ell_{\circlearrowleft}}|3\RD = 0 & \LD6| H_{int,\ell_{\circlearrowleft}}|3\RD = 0 \\
    \hline
     \LD1|H_{int,\ell_{\circlearrowright}}|4\RD =  0 &  \LD2|H_{int,\ell_{\circlearrowright}}|4\RD =0   &  \LD3|H_{int,\ell_{\circlearrowright}}|4\RD = 0.07 & \LD4|H_{int,\ell_{\circlearrowright}}|4\RD =  0& \LD5|H_{int,\ell_{\circlearrowright}}|4\RD =0 & \LD6|H_{int,\ell_{\circlearrowright}}|4\RD =0 \\
    \LD1| H_{int,\ell_{\circlearrowleft}}|4\RD = 0  &  \LD2| H_{int,\ell_{\circlearrowleft}}|4\RD =0.02    &  \LD3| H_{int,\ell_{\circlearrowleft}}|4\RD = 0 & \LD4| H_{int,\ell_{\circlearrowleft}}|4\RD =0  & \LD5| H_{int,\ell_{\circlearrowleft}}|4\RD = 0.21 & \LD6|H_{int,\ell_{\circlearrowleft}}|4\RD =0.07\\
    \hline
     \LD1|H_{int,\ell_{\circlearrowright}}|5\RD =0  & \LD2|H_{int,\ell_{\circlearrowright}}|5\RD =1.85 & \LD3|H_{int,\ell_{\circlearrowright}}|5\RD =0 & \LD4|H_{int,\ell_{\circlearrowright}}|5\RD =0.21 & \LD5|H_{int,\ell_{\circlearrowright}}|5\RD =0& \LD6|H_{int,\ell_{\circlearrowright}}|5\RD =-0.22\\
          \LD1|H_{int,\ell_{\circlearrowleft}}|5\RD =1.22  &   \LD2|H_{int,\ell_{\circlearrowleft}}|5\RD =0&   \LD3|H_{int,\ell_{\circlearrowleft}}|5\RD =0 &   \LD4|H_{int,\ell_{\circlearrowleft}}|5\RD =0 &   \LD5|H_{int,\ell_{\circlearrowleft}}|5\RD =0 &   \LD6|H_{int,\ell_{\circlearrowleft}}|5\RD =0 \\
     \hline
\end{array}$}
\label{tab:SR_TrigoD100}
\end{equation}
%\end{widetext}
\end{minipage}
%\begin{widetext}
 \begin{minipage}{1.0\textwidth}
\begin{equation}
\resizebox{1\hsize}{!}{$
\begin{array}{|c|c|c|c|c|c|}\hline   \Delta_0=200\; meV& & & & &\\
  \hline
  \LD1|H_{int,\ell_{\circlearrowright}}|1\RD = 0 &  \LD2|H_{int,\ell_{\circlearrowright}}|1\RD = -1.16 &  \LD3|H_{int,\ell_{\circlearrowright}}|1\RD =0& \LD4|H_{int,\ell_{\circlearrowright}}|1\RD =0 & \LD5|H_{int,\ell_{\circlearrowright}}|1\RD =0& \LD6|H_{int,\ell_{\circlearrowright}}|1\RD =0.8  \\
    \LD1| H_{int,\ell_{\circlearrowleft}}|1\RD = 0 &  \LD2| H_{int,\ell_{\circlearrowleft}}|1\RD = 0&  \LD3| H_{int,\ell_{\circlearrowleft}}|1\RD =2.18 & \LD4|\ H_{int,\ell_{\circlearrowleft}}|1\RD =0 & \LD5| H_{int,\ell_{\circlearrowleft}}|1\RD =2.17 & \LD6| H_{int,\ell_{\circlearrowleft}}|1\RD =0\\
    \hline
      \LD1|H_{int,\ell_{\circlearrowright}}|2\RD =0&  \LD2|H_{int,\ell_{\circlearrowright}}|2\RD =0   &  \LD3|H_{int,\ell_{\circlearrowright}}|2\RD =2.93  & \LD4|H_{int,\ell_{\circlearrowright}}|2\RD =0 & \LD5|H_{int,\ell_{\circlearrowright}}|2\RD = 0.03 & \LD6|H_{int,\ell_{\circlearrowright}}|2\RD = 0 \\
    \LD1| H_{int,\ell_{\circlearrowleft}}|2\RD = -1.16  &  \LD2| H_{int,\ell_{\circlearrowleft}}|2\RD = 0  &  \LD3| H_{int,\ell_{\circlearrowleft}}|2\RD = 0  & \LD4| H_{int,\ell_{\circlearrowleft}}|2\RD = 0.48& \LD5| H_{int,\ell_{\circlearrowleft}}|2\RD =0 & \LD6| H_{int,\ell_{\circlearrowleft}}|2\RD = 0\\
    \hline
     \LD1|H_{int,\ell_{\circlearrowright}}|3\RD =2.18     &  \LD2|H_{int,\ell_{\circlearrowright}}|3\RD = 0   &  \LD3|H_{int,\ell_{\circlearrowright}}|3\RD =   0& \LD4|H_{int,\ell_{\circlearrowright}}|3\RD =0 & \LD5|H_{int,\ell_{\circlearrowright}}|3\RD =0  & \LD6|H_{int,\ell_{\circlearrowright}}|3\RD =0  \\
    \LD1| H_{int,\ell_{\circlearrowleft}}|3\RD = 0&  \LD2| H_{int,\ell_{\circlearrowleft}}|3\RD = 2.93   &  \LD3| H_{int,\ell_{\circlearrowleft}}|3\RD =  0  & \LD4| H_{int,\ell_{\circlearrowleft}}|3\RD =  0.90 & \LD5| H_{int,\ell_{\circlearrowleft}}|3\RD = 0 & \LD6| H_{int,\ell_{\circlearrowleft}}|3\RD = -2.02 \\
    \hline
     \LD1|H_{int,\ell_{\circlearrowright}}|4\RD =  0 &  \LD2|H_{int,\ell_{\circlearrowright}}|4\RD =0.48&  \LD3|H_{int,\ell_{\circlearrowright}}|4\RD = 0.90 & \LD4|H_{int,\ell_{\circlearrowright}}|4\RD =  0& \LD5|H_{int,\ell_{\circlearrowright}}|4\RD =0 &  \LD6|H_{int,\ell_{\circlearrowright}}|4\RD = 0.33\\
    \LD1| H_{int,\ell_{\circlearrowleft}}|4\RD = 0  &  \LD2| H_{int,\ell_{\circlearrowleft}}|4\RD =0    &  \LD3| H_{int,\ell_{\circlearrowleft}}|4\RD = 0 & \LD4| H_{int,\ell_{\circlearrowleft}}|4\RD =0  & \LD5| H_{int,\ell_{\circlearrowleft}}|4\RD = 0.90  & \LD6| H_{int,\ell_{\circlearrowleft}}|4\RD =0 \\
    \hline
     \LD1|H_{int,\ell_{\circlearrowright}}|5\RD =  2.17 & \LD2 |H_{int,\ell_{\circlearrowright}}|5\RD = 0 &  \LD3 |H_{int,\ell_{\circlearrowright}}|5\RD =  0 &  \LD4 |H_{int,\ell_{\circlearrowright}}|5\RD =  0.90 &  \LD5 |H_{int,\ell_{\circlearrowright}}|5\RD =  0  &  \LD6|H_{int,\ell_{\circlearrowright}}|5\RD = 0 \\  
      \LD1|H_{int,\ell_{\circlearrowleft}}|5\RD = 0 &    \LD2|H_{int,\ell_{\circlearrowleft}}|5\RD = 0.03 &   \LD3|H_{int,\ell_{\circlearrowleft}}|5\RD = 0 &   \LD4|H_{int,\ell_{\circlearrowleft}}|5\RD = 0 &   \LD5|H_{int,\ell_{\circlearrowleft}}|5\RD = 0  &   \LD6|H_{int,\ell_{\circlearrowleft}}|5\RD = -0.02 \\  
      \hline 
       \LD7|H_{int,\ell_{\circlearrowright}}|1\RD = -1.50 &   \LD8|H_{int,\ell_{\circlearrowright}}|1\RD = 0 &   \LD9|H_{int,\ell_{\circlearrowright}}|1\RD = 0 &       \LD7|H_{int,\ell_{\circlearrowright}}|2\RD = 0 &   \LD8|H_{int,\ell_{\circlearrowright}}|2\RD = -1.15 &   \LD9|H_{int,\ell_{\circlearrowright}}|2\RD = 0 \\
              \LD7|H_{int,\ell_{\circlearrowleft}}|1\RD = 0 &   \LD8|H_{int,\ell_{\circlearrowleft}}|1\RD = -0.93 &   \LD9|H_{int,\ell_{\circlearrowleft}}|1\RD = 0 &  \LD7|H_{int,\ell_{\circlearrowleft}}|2\RD = 0 &   \LD8|H_{int,\ell_{\circlearrowleft}}|2\RD = 0 &   \LD9|H_{int,\ell_{\circlearrowleft}}|2\RD = 0\\
      \hline
             \LD7|H_{int,\ell_{\circlearrowright}}|3\RD = 0 &   \LD8|H_{int,\ell_{\circlearrowright}}|3\RD = 0 &   \LD9|H_{int,\ell_{\circlearrowright}}|3\RD = 0 &       \LD7|H_{int,\ell_{\circlearrowright}}|4\RD = 0.62 &   \LD8|H_{int,\ell_{\circlearrowright}}|4\RD = 0 &   \LD9|H_{int,\ell_{\circlearrowright}}|4\RD = 0 \\
              \LD7|H_{int,\ell_{\circlearrowleft}}|3\RD = 3.80 &   \LD8|H_{int,\ell_{\circlearrowleft}}|3\RD = 0 &   \LD9|H_{int,\ell_{\circlearrowleft}}|3\RD = 0 &  \LD7|H_{int,\ell_{\circlearrowleft}}|4\RD = 0 &   \LD8|H_{int,\ell_{\circlearrowleft}}|4\RD = 0.38 &   \LD9|H_{int,\ell_{\circlearrowleft}}|4\RD = 0\\
              \hline
                \LD7|H_{int,\ell_{\circlearrowright}}|5\RD = 0 &   \LD8|H_{int,\ell_{\circlearrowright}}|5\RD = 0 &   \LD9|H_{int,\ell_{\circlearrowright}}|5\RD = 0 &       \LD7|H_{int,\ell_{\circlearrowright}}|6\RD = 0 &   \LD8|H_{int,\ell_{\circlearrowright}}|6\RD = 0.79 &   \LD9|H_{int,\ell_{\circlearrowright}}|6\RD = 0 \\
              \LD7|H_{int,\ell_{\circlearrowleft}}|5\RD = 0.04 &   \LD8|H_{int,\ell_{\circlearrowleft}}|5\RD = 0 &   \LD9|H_{int,\ell_{\circlearrowleft}}|5\RD = 0 &  \LD7|H_{int,\ell_{\circlearrowleft}}|6\RD = 0 &   \LD8|H_{int,\ell_{\circlearrowleft}}|6\RD = 0 &   \LD9|H_{int,\ell_{\circlearrowleft}}|6\RD = 0\\
      \hline
       \LD7|H_{int,\ell_{\circlearrowright}}|7\RD = 0 &  \LD8|H_{int,\ell_{\circlearrowright}}|7\RD = -1.49 &  \LD9|H_{int,\ell_{\circlearrowright}}|7\RD = 0&            \LD7|H_{int,\ell_{\circlearrowright}}|8\RD = 0 &  \LD8|H_{int,\ell_{\circlearrowright}}|8\RD = 0&  \LD9|H_{int,\ell_{\circlearrowright}}|8\RD =0\\
              \LD7|H_{int,\ell_{\circlearrowleft}}|7\RD = 0 &  \LD8|H_{int,\ell_{\circlearrowleft}}|7\RD = 0&  \LD9|H_{int,\ell_{\circlearrowleft}}|7\RD =0 &          \LD7|H_{int,\ell_{\circlearrowleft}}|8\RD =  -1.49 &  \LD8|H_{int,\ell_{\circlearrowleft}}|8\RD = 0&  \LD9|H_{int,\ell_{\circlearrowleft}}|8\RD =0 \\
              \hline
\end{array}$}
\label{tab:SR_TrigoD200}
\end{equation}
Transition matrix elements $t^+_{n^{\prime},n} $  (in units of $\frac{meV}{E_{\omega}[mV/nm]/\hbar\omega[meV]} $) quantifying the absorption of right ($\boldsymbol{\ell}_{\circlearrowright}$) and left-handed ($\boldsymbol{\ell}_{\circlearrowleft}$) circularly polarized light between the states of dots with different gaps, $\Delta_0$.
\end{minipage}
%\end{widetext} 
\clearpage

\section{Numerical Diagonalisation of the Two-Particle Hamiltonian}

We use the confined single-particle wave functions of the $n$th dot level $\SPstate_{n}(\boldsymbol{r})=\sum_{(\eta ,\mu)}a_{(\eta ,\mu)_{n}}\SPWF_{\eta}(x)\SPWF_{\mu}(y)$ from the numerics in the basis of HO eigenfunctions. This allows us to compute the  matrix element of the SU(4)-symmetric Coulomb interaction as 
\begin{widetext}
\begin{align}
\nonumber &V_{\substack{n_1,n_2\\n_3,n_4}} = \iint d\mathbf{r}_1 d\mathbf{r}_2 \; \SPstate_{n_1}(\boldsymbol{r}_1)^*  \SPstate_{n_2}(\boldsymbol{r}_1) V(\boldsymbol{r}_1-\boldsymbol{r}_2) \SPstate_{n_3}(\boldsymbol{r}_2)^*  \SPstate_{n_4}(\boldsymbol{r}_2)\\ 
\nonumber&= \int d\mathbf{q} V(\mathbf{q}) \sum_{(\eta  \mu)_{n_{1,2,3,4}}} a_{(\eta  ,\mu)_{n_1}}^* a_{(\eta  ,\mu)_{n_2}} a_{(\eta  ,\mu)_{n_3}}^* a_{(\eta  ,\mu)_{n_4}} \\
\nonumber&\underbrace{ \int dx_1\SPWF_{\eta _1} (x_1) \SPWF_{\eta _2} (x_1) e^{iq_x x_1} }_{F_{\eta _1,\eta _2}(q_x)} \underbrace{ \int dy_1\SPWF_{\mu_1} (y_1) \SPWF_{\mu_2} (y_1) e^{iq_y y_1} }_{F_{\mu_1,\mu_2}(q_y)} \underbrace{\int dx_2\SPWF_{\eta _3} (x_2) \SPWF_{\eta _4} (x_2) e^{-iq_x x_2}}_{F_{\eta _3,\eta _4}(-q_x)} \underbrace{ \int dy_2\SPWF_{\mu_3} (y_2) \SPWF_{\mu_4} (y_2) e^{-iq_y y_2} }_{F_{\mu_3,\mu_4}(-q_y)} \\
&= \int d\mathbf{q} V(\mathbf{q}) \sum_{(\eta  \mu)_{n_{1,2,3,4}}} a_{(\eta ,\mu)_{n_1}}^* a_{(\eta ,\mu)_{n_2}} a_{(\eta ,\mu)_{n_3}}^* a_{(\eta ,\mu)_{n_4}} F_{\eta _1,\eta _2}(q_x) F_{\mu_1,\mu_2}(q_y)F_{\eta _3,\eta _4}(-q_x) F_{\mu_3,\mu_4}(-q_y).
\end{align}
\end{widetext}
Here, we exploited that the plane-wave matrix elements of harmonic oscillator eigenfunctions can be obtained analytically and read
\begin{equation}
F_{\eta _1,\eta _2}({q})=\sqrt{\frac{\eta _2 !}{\eta _1 !}} \Big(\frac{ i q}{\sqrt{2}\alpha}\Big)^{\eta_1-\eta_2} e^{-\frac{q^2}{4\alpha^2}} L_{\eta_2}^{\eta _1-\eta _2}[\frac{q^2}{2\alpha^2}],
\end{equation}
for $\eta_2\le \eta _1 $ and $F_{\eta _2,\eta _1}( {q})=[F_{\eta _1,\eta _2}(- {q})]^*$. Further, $ L_{\eta }^{\mu}(x)$ denotes the generalized Laguerre polynomial.\\

\begin{figure}[t!]
 \centering
\includegraphics[width=0.8\linewidth]{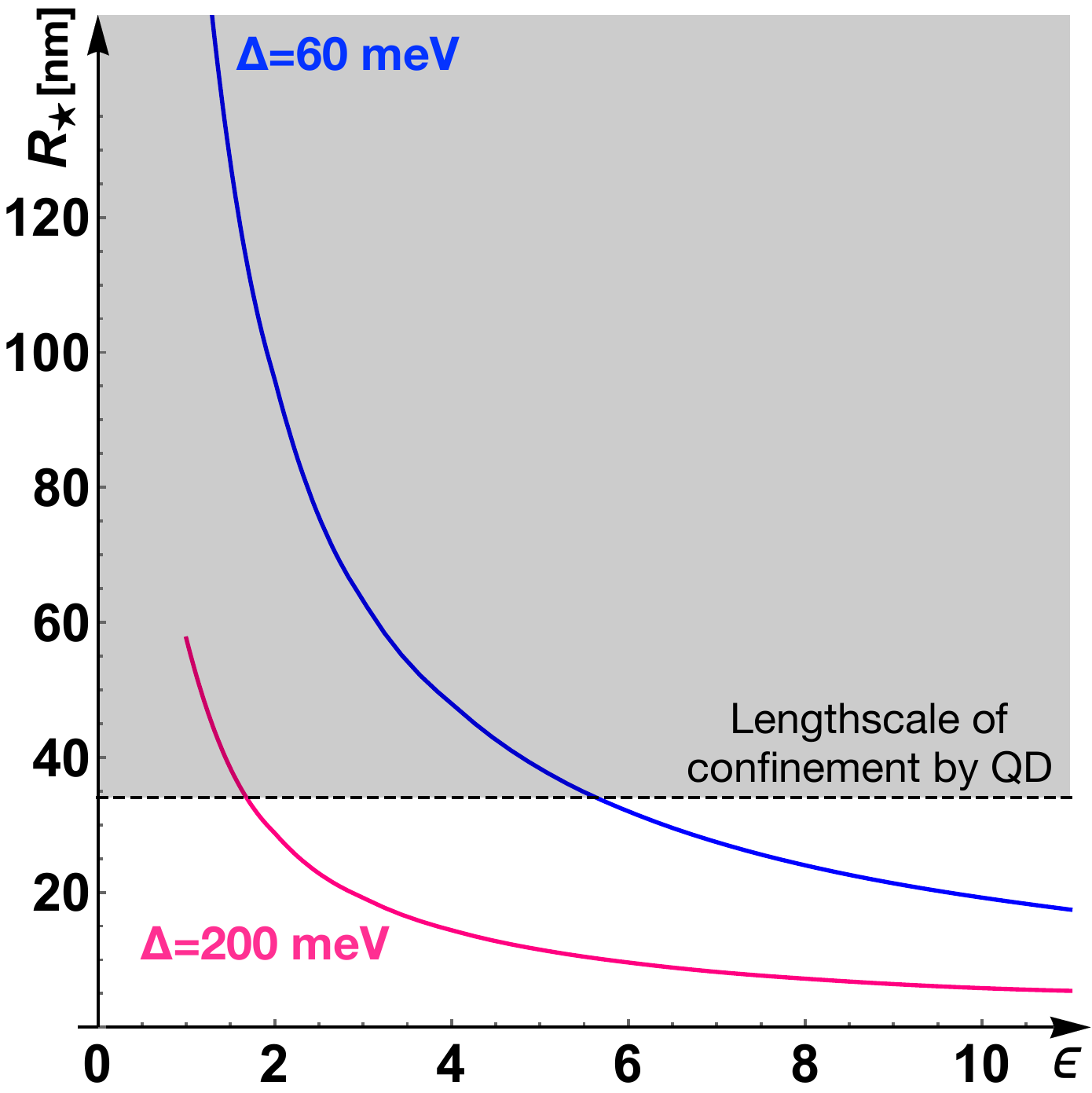}
\caption{Screening length, $R_{\star}$, for a small gap, $\Delta_0=60$ meV, and a large gap, $\Delta_0=200$ meV, as a function of $\epsilon$. The dashed line compares to lengthscale of confinement by the QD, obtained from the width of the probability distributions of the confined wave functions.}
\label{fig:QD_R}
\end{figure}

%\begin{figure}[t!]
 %\centering
%\includegraphics[width=0.7\linewidth]{QD_XD}
%\caption{ Exchange and direct matrix elements $X_{12}$ and $D_{12}$ between the two lowest \SP levels as functions of $\epsilon$.}
%\label{fig:QD_XD}
%\end{figure}
We chose the interaction kernel $V(\mathbf{q})$  as that of the  2D screened Coulomb interaction in a weakly gapped BLG ,
\begin{equation}
V(\mathbf{q})=\frac{e^2}{4\pi\epsilon_0\epsilon}\frac{2\pi}{q(1+q R_{\star})}.
\end{equation}
 The screening length $R_{\star}=  \sqrt{32} \hbar\kappa /\sqrt{\Delta}$ (with polarizability $\kappa^2 =2m e^2/(4\pi\epsilon_0\epsilon \hbar \sqrt{\Delta} )^2  $) defines the spatial range in which the interaction is a Keldysh potential (for $r< R_{\star}$), or exhibits $1/r$ behaviour (for $r\gg R_{\star}$). Among others, $R_{\star}$ depends on the gap induced in the BLG as well as on the dielectric constant, $\epsilon$, of the encapsulating material.  In \fig\ref{fig:QD_R} we show $R_{\star}$ for a small gap, $\Delta_0=60$ meV, and a large gap, $\Delta_0=200$ meV, as a function of $\epsilon$. In this figure, we also compare the screening length to the lengthscale of confinement by the QD, which we define as the width of the probability distributions of the confined wave functions. 

\section{Estimation of short-range couplings}

To estimate the contribution of the short range symmetry breaking interaction we start from the form  \cite{Lemonik2010, Lemonik2012, Aleiner2007},
\begin{equation}
H_{SR}=\frac{1}{2}\iint d^2 \mathbf{r}^2\sum_{ij} g_{ij}\big[ \Upsilon^{\dagger}(\mathbf{r})\sigma_i\tau_j  \Upsilon (\mathbf{r}) \big]^2, 
\label{eqn:HSR}
\end{equation}
where $\sigma_i$ $(\tau_i)$ are the Pauli matrices acting in sub-layer (valley) space. %and $T=\text{diag}(1,1,1,-1)$.
The state $\Upsilon$ comprises the low-energy states on the non-dimer sites $A, B'$ of the BLG lattice assembled in the spinor $\Upsilon=\{ \psi^K_{A}, \psi^K_{B'},\psi^{K'}_{B'}    ,-\psi^{K'}_{A}\}$.

Writing the Hamiltonian of \eqn\ref{eqn:HSR} in matrix form,
\begin{equation}
H_{SR}=\frac{1}{2}\begin{pmatrix}
 H_{\substack{K^+K^+\\K^+K^+}} & H_{\substack{K^+K^-\\K^+K^+}} & H_{\substack{K^-K^+\\K^+K^+}}&H_{\substack{K^-K^-\\K^+K^+}}\\
 H_{\substack{K^+K^+\\K^+K^-}} & H_{\substack{K^+K^-\\K^+K^-}} &H_{\substack{K^+K^-\\K^-K^+}} &H_{\substack{K^-K^-\\K^+K^-}} \\
H_{\substack{K^+K^+\\K^-K^+}}& H_{\substack{K^+K^-\\K^-K^+}}& H_{\substack{K^-K^+\\K^-K^+}}& H_{\substack{K^-K^-\\K^-K^+}}\\
H_{\substack{K^+K^+\\K^-K^-}}& H_{\substack{K^+K^-\\K^-K^-}}& H_{\substack{K^-K^+\\K^-K^-}}& H_{\substack{K^-K^-\\K^-K^-}},
\end{pmatrix},
\label{eqn:HamSR}
\end{equation}
%and evaluate the individual matrix elements using the numerical SP wave-functions of the GS inside the dot:
%\begin{equation}
%H_{\substack{\xi_1 \xi_2\\ \xi_3 \xi_4}} =\int d \boldsymbol{r} \; [\psi^{B'}_{\xi_3}(\boldsymbol{r} )]^*  \psi^{B'}_{ \xi_1}(\boldsymbol{r} )  [\psi^{B'}_{\xi_4}(\boldsymbol{r} )]^*  \psi^{B'}_{ \xi_2}(\boldsymbol{r} ) .
%\end{equation}
we use the  envelope wave functions $\SPstate$ of the lowest \SP dot level at $\Delta_0=60$ meV ($\Delta_0=200$ meV)  and $L=80$ nm to compute the  prefactors   numerically. When evaluating the prefactors for the lowest symmetric dot state
\begin{equation}
W=W_{11,11}=\frac{1}{2}\int d\mathbf{r}\,\TPorbWF^s\,^*_{11}(\mathbf{r}) \TPorbWF^s_{11}(\mathbf{r}),
\label{eqn:W}
\end{equation}
due to layer polarization in BLG with an external displacement field, we find  find non-zero matrix elements with weights
\begin{align}
W=\nonumber &\iint d \boldsymbol{r} \; [\SPstate^{B'}_{K^+}(\boldsymbol{r} )]^*  \SPstate^{B'}_{K^+}(\boldsymbol{r} )  [\SPstate^{B'}_{K^+}(\boldsymbol{r} )]^*  \SPstate^{B'}_{K^+}(\boldsymbol{r} ) \\
\nonumber&=\iint d \boldsymbol{r} \; [\SPstate^{B'}_{K^+}(\boldsymbol{r} )]^*  \SPstate^{B'}_{K^+}(\boldsymbol{r} )  [\SPstate^{B'}_{K^-}(\boldsymbol{r} )]^*  \SPstate^{B'}_{K^-}(\boldsymbol{r} ) \\
\nonumber&=\iint d \boldsymbol{r} \; [\SPstate^{B'}_{K^-}(\boldsymbol{r} )]^*  \SPstate^{B'}_{K^+}(\boldsymbol{r} )  [\SPstate^{B'}_{K^+}(\boldsymbol{r} )]^*  \SPstate^{B'}_{K^-}(\boldsymbol{r} ) \\
\nonumber&=\iint d \boldsymbol{r} \; [\SPstate^{B'}_{K^-}(\boldsymbol{r} )]^*  \SPstate^{B'}_{K^+}(\boldsymbol{r} )  [\SPstate^{B'}_{K^-}(\boldsymbol{r} )]^*  \SPstate^{B'}_{K^+}(\boldsymbol{r} ) \\
&=\frac{4.3*10^{-4}}{nm^2} \Big(\frac{3.7*10^{-4}}{nm^2}\Big),
\end{align}
as well as their complex conjugates, while all other matrix elements in \eqn\eqref{eqn:HamSR} vanish. Hence, the Hamiltonian in \eqn\eqref{eqn:HamSR} reduces to
\begin{widetext}
\begin{equation}H_{SR}={W}
\begin{pmatrix}
  g_{z0} + g_{0z}+ g_{zz} & 0 & 0&0\\
 0 &  -g_{z0} - g_{0z}+ g_{zz}&g_{xx}+g_{yy}+g_{xy}+g_{yx} &0 \\
0& g_{xx}+g_{yy}+g_{xy}+g_{yx} &   -g_{z0} - g_{0z}+ g_{zz}& 0\\
0& 0& 0&  g_{z0} + g_{0z}+ g_{zz}
\end{pmatrix}.
\label{eqn:HamSRred}
\end{equation}
\end{widetext}

 We estimate corresponding couplings $g_{ij}$ in a simple tight-binding model for graphene's $p_z$-orbital electrons\cite{Danovich2017}.  Writing the Bloch part   of the wave function  as $\upsilon(\mathbf{r})=(u^{A}_{ K^{+}}(\mathbf{r}),u^{B^{\prime}}_{ K^{+}}(\mathbf{r}),u^{B^{\prime}}_{ K^{-}}(\mathbf{r}),-u^{A}_{ K^{-}}(\mathbf{r}))^T$, the couplings read  \cite{Lemonik2010, Lemonik2012, Aleiner2007, Kharitonov2012a}
\begin{equation}
g_{ij}\propto\int d^3  \boldsymbol{r}_1\int_{\substack{unit\\cell}}  d \boldsymbol{r}^3 _2 \frac{\rho_{ij}( \boldsymbol{r}_1)\rho_{ij}( \boldsymbol{r}_2)}{A|\boldsymbol{r}_1-\boldsymbol{r}_2|},
\end{equation}
where $A=|\mathbf{a}_1\times \mathbf{a}_2|$ is the unit cell area of graphene computed from graphene's real space lattice vectors $\mathbf{a}_1, \mathbf{a}_2$, and 
\begin{equation}
\rho_{ij}( \boldsymbol{r})= \frac{1}{2} \upsilon^{\dagger}( \boldsymbol{r})\sigma_i \tau_j  \upsilon( \boldsymbol{r}).
\end{equation}
For the Bloch functions we choose the form 
 \begin{align}
\nonumber u^{A_1/B_2}_{K^{\pm}} (\mathbf{r})&=\frac{\sqrt{A}}{\sqrt{\mathcal{N}}} \sum_{\mathbf{R}_i}e^{iK_{\pm}\mathbf{R}_i} \varphi_{210}(\mathbf{r}-\mathbf{R}_i),\\
\varphi_{210}(\mathbf{r})&=\frac{N}{\sqrt{a_0^3}}\,P( {r})Y^1_0(\theta,\phi),
\end{align}
ensuring normalization \cite{Callaway1991} $\int_{\substack{unit\\cell}} u^{*}(\boldsymbol{r}) u(\boldsymbol{r})\,d^3\mathbf{r} = A.$
In the equations above, $N=\frac{2Z}{\sqrt{32}}\sqrt{\frac{Z^3}{3}}$, $P(r)=\frac{r}{a_0}e^{-\frac{Z}{2}\frac{r}{a_0}}$, and $Y^l_m$ denote the spherical harmonics of graphene's 2$p_z$ state. Further, Z is the atomic number,  $a_0=\frac{4\pi \epsilon_0\hbar}{m_e e^2 Z}$ is the Bohr radius, and $\{\mathbf{R}_i\}$ denote the lattice points of the sublattice the wave function is residing on. 
%\clearpage
We distinguish between the different scattering processes,
\begin{widetext}
 \begin{align}
 \nonumber M_{\substack{intra\\same}} &=\frac{e^2}{4\pi\epsilon_C\epsilon_0}\int d\mathbf{r}_1^3\int_{\substack{unit\\cell}}  d\mathbf{r}_2^3  \frac{1}{4A|\mathbf{r}_2-\mathbf{r}_1|} (u^{A_1}_{K^{+}})^* (\mathbf{r}_1)  (u^{A_1}_{K^{+}})^* (\mathbf{r}_2)  u^{A_1}_{K^{+}}  (\mathbf{r}_1)  u^{A_1}_{K^{+}}  (\mathbf{r}_2) \\
\nonumber  &=\frac{e^2}{4\pi\epsilon_C\epsilon_0}\int d\mathbf{r}_1^3\int_{\substack{unit\\cell}}  d\mathbf{r}_2^3  \frac{1}{4A|\mathbf{r}_2-\mathbf{r}_1|} (u^{A_1}_{K^{-}})^* (\mathbf{r}_1)  (u^{A_1}_{K^{-}})^* (\mathbf{r}_2)  u^{A_1}_{K^{-}}  (\mathbf{r}_1)  u^{A_1}_{K^{-}}  (\mathbf{r}_2) \\
\nonumber  &=\frac{e^2}{4\pi\epsilon_C\epsilon_0}\int d\mathbf{r}_1^3\int_{\substack{unit\\cell}}  d\mathbf{r}_2^3  \frac{1}{4A|\mathbf{r}_2-\mathbf{r}_1|} (u^{B_2}_{K^{+}})^* (\mathbf{r}_1)  (u^{B_2}_{K^{+}})^* (\mathbf{r}_2)  u^{B_2}_{K^{+}}  (\mathbf{r}_1)  u^{B_2}_{K^{+}}  (\mathbf{r}_2) \\
\nonumber&=\frac{e^2}{4\pi\epsilon_C\epsilon_0}\int d\mathbf{r}_1^3\int_{\substack{unit\\cell}}  d\mathbf{r}_2^3  \frac{1}{4A|\mathbf{r}_2-\mathbf{r}_1|} (u^{B_2}_{K^{-}})^* (\mathbf{r}_1)  (u^{B_2}_{K^{-}})^* (\mathbf{r}_2)  u^{B_2}_{K^{-}}  (\mathbf{r}_1)  u^{B_2}_{K^{-}}  (\mathbf{r}_2) \\
 \nonumber M_{\substack{intra\\diff}} &=\frac{e^2}{4\pi\epsilon_C\epsilon_0}\int d\mathbf{r}_1^3\int_{\substack{unit\\cell}}  d\mathbf{r}_2^3  \frac{1}{4A|\mathbf{r}_2-\mathbf{r}_1|} (u^{A_1}_{K^{+}})^* (\mathbf{r}_1)  (u^{B_2}_{K^{+}})^* (\mathbf{r}_2)  u^{A_1}_{K^{+}}  (\mathbf{r}_1)  u^{B_2}_{K^{+}}  (\mathbf{r}_2) \\
 \nonumber&= \frac{e^2}{4\pi\epsilon_C\epsilon_0}\int d\mathbf{r}_1^3\int_{\substack{unit\\cell}}  d\mathbf{r}_2^3  \frac{1}{4A|\mathbf{r}_2-\mathbf{r}_1|} (u^{A_1}_{K^{-}})^* (\mathbf{r}_1)  (u^{B_2}_{K^{-}})^* (\mathbf{r}_2)  u^{A_1}_{K^{-}}  (\mathbf{r}_1)  u^{B_2}_{K^{-}}  (\mathbf{r}_2) \\
 \nonumber &=\frac{e^2}{4\pi\epsilon_C\epsilon_0}\int d\mathbf{r}_1^3\int_{\substack{unit\\cell}}  d\mathbf{r}_2^3  \frac{1}{4A|\mathbf{r}_2-\mathbf{r}_1|} (u^{B_2}_{K^{+}})^* (\mathbf{r}_1)  (u^{A_1}_{K^{+}})^* (\mathbf{r}_2)  u^{B_2}_{K^{+}}  (\mathbf{r}_1)  u^{A_1}_{K^{+}}  (\mathbf{r}_2) \\
  \nonumber &=\frac{e^2}{4\pi\epsilon_C\epsilon_0}\int d\mathbf{r}_1^3\int_{\substack{unit\\cell}}  d\mathbf{r}_2^3  \frac{1}{4A|\mathbf{r}_2-\mathbf{r}_1|} (u^{B_2}_{K^{-}})^* (\mathbf{r}_1)  (u^{A_1}_{K^{-}})^* (\mathbf{r}_2)  u^{B_2}_{K^{-}}  (\mathbf{r}_1)  u^{A_1}_{K^{-}}  (\mathbf{r}_2) \\
 \nonumber M_{\substack{inter\\same}} &=\frac{e^2}{4\pi\epsilon_C\epsilon_0}\int d\mathbf{r}_1^3\int_{\substack{unit\\cell}}  d\mathbf{r}_2^3  \frac{1}{4A|\mathbf{r}_2-\mathbf{r}_1|} (u^{A_1}_{K^{-}})^* (\mathbf{r}_1)  (u^{A_1}_{K^{+}})^* (\mathbf{r}_2)  u^{A_1}_{K^{+}}  (\mathbf{r}_1)  u^{A_1}_{K^{-}}  (\mathbf{r}_2) \\
 \nonumber &=\frac{e^2}{4\pi\epsilon_C\epsilon_0}\int d\mathbf{r}_1^3\int_{\substack{unit\\cell}}  d\mathbf{r}_2^3  \frac{1}{4A|\mathbf{r}_2-\mathbf{r}_1|} (u^{A_1}_{K^{+}})^* (\mathbf{r}_1)  (u^{A_1}_{K^{-}})^* (\mathbf{r}_2)  u^{A_1}_{K^{-}}  (\mathbf{r}_1)  u^{A_1}_{K^{+}}  (\mathbf{r}_2)\\
\nonumber &=\frac{e^2}{4\pi\epsilon_C\epsilon_0}\int d\mathbf{r}_1^3\int_{\substack{unit\\cell}}  d\mathbf{r}_2^3  \frac{1}{4A|\mathbf{r}_2-\mathbf{r}_1|} (u^{B_2}_{K^{-}})^* (\mathbf{r}_1)  (u^{B_2}_{K^{+}})^* (\mathbf{r}_2)  u^{B_2}_{K^{+}}  (\mathbf{r}_1)  u^{B_2}_{K^{-}}  (\mathbf{r}_2) \\
\nonumber&=\frac{e^2}{4\pi\epsilon_C\epsilon_0}\int d\mathbf{r}_1^3\int_{\substack{unit\\cell}}  d\mathbf{r}_2^3  \frac{1}{4A|\mathbf{r}_2-\mathbf{r}_1|} (u^{B_2}_{K^{+}})^* (\mathbf{r}_1)  (u^{B_2}_{K^{-}})^* (\mathbf{r}_2)  u^{B_2}_{K^{-}}  (\mathbf{r}_1)  u^{B_2}_{K^{+}}  (\mathbf{r}_2)\\
 \nonumber M_{\substack{inter\\diff}} &=\frac{e^2}{4\pi\epsilon_C\epsilon_0}\int d\mathbf{r}_1^3\int_{\substack{unit\\cell}}  d\mathbf{r}_2^3  \frac{1}{4A|\mathbf{r}_2-\mathbf{r}_1|} (u^{A_1}_{K^{-}})^* (\mathbf{r}_1)  (u^{B_2}_{K^{+}})^* (\mathbf{r}_2)  u^{A_1}_{K^{+}}  (\mathbf{r}_1)  u^{B_2}_{K^{-}}  (\mathbf{r}_2) \\
\nonumber &=\frac{e^2}{4\pi\epsilon_C\epsilon_0}\int d\mathbf{r}_1^3\int_{\substack{unit\\cell}}  d\mathbf{r}_2^3  \frac{1}{4A|\mathbf{r}_2-\mathbf{r}_1|} (u^{A_1}_{K^{+}})^* (\mathbf{r}_1)  (u^{B_2}_{K^{-}})^* (\mathbf{r}_2)  u^{A_1}_{K^{-}}  (\mathbf{r}_1)  u^{B_2}_{K^{+}}  (\mathbf{r}_2) \\
\nonumber &=\frac{e^2}{4\pi\epsilon_C\epsilon_0}\int d\mathbf{r}_1^3\int_{\substack{unit\\cell}}  d\mathbf{r}_2^3  \frac{1}{4A|\mathbf{r}_2-\mathbf{r}_1|} (u^{B_2}_{K^{-}})^* (\mathbf{r}_1)  (u^{A_1}_{K^{+}})^* (\mathbf{r}_2)  u^{B_2}_{K^{+}}  (\mathbf{r}_1)  u^{A_1}_{K^{-}}  (\mathbf{r}_2) \\
&=\frac{e^2}{4\pi\epsilon_C\epsilon_0}\int d\mathbf{r}_1^3\int_{\substack{unit\\cell}}  d\mathbf{r}_2^3  \frac{1}{4A|\mathbf{r}_2-\mathbf{r}_1|} (u^{B_2}_{K^{+}})^* (\mathbf{r}_1)  (u^{A_1}_{K^{-}})^* (\mathbf{r}_2)  u^{B_2}_{K^{-}}  (\mathbf{r}_1)  u^{A_1}_{K^{+}}  (\mathbf{r}_2) ,
 \end{align}
 \end{widetext}
where "intra/inter" refer to inter- and intra-valley scattering and "same/different" label processes on the same sublattice or involving different sublattices, respectively.

Within a two-centre approximation we can write the real space integral as 
\begin{widetext}
 \begin{align}
\nonumber M&=\frac{e^2}{4\pi\epsilon_C\epsilon_0}\int d\mathbf{r}_1^3\int_{\substack{unit\\cell}}  d\mathbf{r}_2^3  \frac{1}{4A|\mathbf{r}_2-\mathbf{r}_1|} (u^{A_1/B_2}_{K^{\pm}})^* (\mathbf{r}_1)  (u^{A_1/B_2}_{K^{\pm}})^* (\mathbf{r}_2)  u^{A_1/B_2}_{K^{\pm}}  (\mathbf{r}_1)  u^{A_1/B_2}_{K^{\pm}}  (\mathbf{r}_2)  \\
&\approx \frac{e^2}{4\pi\epsilon_C\epsilon_0} \frac{A}{4} \sum_{\mathbf{R}}e^{i\Delta \mathbf{K}\cdot \mathbf{R}} \iint d\mathbf{r}_1^3 d\mathbf{r}_2^3  \frac{1}{|\mathbf{r}_2-\mathbf{r}_1+\mathbf{R}|} |\varphi_{210}(\mathbf{r}_1)|^2 |\varphi_{210}(\mathbf{r}_2)|^2,
\label{eqn:M}
\end{align}
\end{widetext}
where $\mathbf{R}=\mathbf{R}^{(2)}-\mathbf{R}^{(1)}$ connects the centres of the two wave packages.

We evaluate the real-space integral in \eqn\eqref{eqn:M} in two steps. For $\mathbf{R}\equiv0$ we can use the Laplace expansion of the Coulomb interaction,
 \begin{equation}
 \frac{1}{|\mathbf{r}_2-\mathbf{r}_1 |}  =  \sum_{l=0}^{\infty}\frac{r_{<}^l}{r_{>}^{l+1}}\sum_{m=-l}^{m=+l}\frac{4\pi}{2l+1}(Y_m^l)^*(\theta_1,\phi_1)Y_m^l (\theta_2,\phi_2),
\end{equation}
and the identity 
\begin{align} 
\nonumber & \int \sin\theta d\theta d\phi Y_{m_1}^{l_1} (\theta,\phi)  Y_{m_2}^{l_2} (\theta,\phi)  Y_{m_3}^{l_3} (\theta,\phi) \\
&= \sqrt{\frac{(2l_1+1)(2l_2+1)(2l_2+1)}{4\pi}}\begin{pmatrix}
 l_1 & l_2 & l_3\\
 0 & 0 &0
\end{pmatrix}
\begin{pmatrix}
 l_1 & l_2 & l_3\\
 m_1 & m_2 &m_3
\end{pmatrix},
\end{align}
and the integral above evaluates to 
\begin{align}
\nonumber&\mathcal{M}_{0}\\
\nonumber&= \frac{e^2}{4\pi\epsilon_C\epsilon_0} \frac{A}{4} \iint d\mathbf{r}_1^3 d\mathbf{r}_2^3 \frac{1}{|\mathbf{r}_2-\mathbf{r}_1 |}  |\varphi_{210}(\mathbf{r}_1)|^2  |\varphi_{210}(\mathbf{r}_2)|^2\\
&=\frac{e^2}{4\pi\epsilon_C\epsilon_0 a_0}  \frac{A}{4} \,Z\frac{501}{2560}.
\label{eqn:Res_R0}
\end{align}

In the case $\mathbf{R}\neq0$ we employ the expansion\cite{Arrighini1981, Solovyov2007} for $|\mathbf{r}_2-\mathbf{r}_1 |<R$,
 \begin{align}
  \frac{1}{|\mathbf{r}_2-\mathbf{r}_1+\mathbf{R}|}=\sum_{l_a,l_b=0}^{\infty}R^{-(l_a+l_b+1)}r_1^{l_a} r_2^{l_b} V_{l_a,l_b};
 \end{align} 
 where 
 \begin{widetext}
\begin{align}
\nonumber& V_{l_a,l_b}= (4\pi)^{\frac{3}{2}}(-1)^{l_b}\begin{pmatrix}
 2(l_a+l_b)\\
 2l_a
\end{pmatrix}^{\frac{1}{2}} \frac{1}{\sqrt{(2l_a+1)(2l_b+1)(2(l_a+l_b)+1)}}\\
&\times \sum_{M=-(l_a+l_b)}^{l_a+l_b}(-1)^M Y_{-M}^L(\theta_R,\phi_R)\bigg( \sum_{m_a=-l_a}^{l_a}\sum_{m_b=-l_b}^{l_b} Y_{m_a}^{l_a}(\theta_1,\phi_1)Y_{m_b}^{l_b}(\theta_2,\phi_2)\langle l_a m_a;l_b m_b | (l_a+l_b)M\rangle\bigg).
\end{align}
 \end{widetext}
This allows as to compute the integral according to 
\begin{align}
 \nonumber&\mathcal{M}_{\mathbf{R}\neq0}\\
 \nonumber&= \frac{e^2}{4\pi\epsilon_C\epsilon_0} \frac{A}{4}\iint d\mathbf{r}_1^3 d\mathbf{r}_2^3 \frac{1}{|\mathbf{r}_2-\mathbf{r}_1+\mathbf{R} |}  |\Psi_{210}(\mathbf{r}_1)|^2  |\Psi_{210}(\mathbf{r}_2)|^2 \\
 \nonumber &=\frac{e^2}{4\pi\epsilon_C\epsilon_0 a_0} \frac{A}{4}\big[  \frac{1}{(R/a_0)}  +\frac{2}{25\sqrt{\pi}}\frac{1}{(R/a_0)^3}Y_0^2(\theta_R,\phi_R)\frac{30}{Z^{2}}   \\
 &+ \frac{4}{125\sqrt{\pi}}\frac{1}{(R/a_0)^5}Y_0^4(\theta_R,\phi_R)\frac{900}{Z^{4}}     \big].
 \label{eqn:Res_Rneq0}
 \end{align} 

Using \eqns\eqref{eqn:Res_R0} and \eqref{eqn:Res_Rneq0} the sum over the vectors $\{\mathbf{R}\}$ in \eqn\eqref{eqn:M} can be carried out numerically. 

We calculate the couplings involving the different scattering processes above,
\begin{align}
\nonumber g_{0z}&=g_{z0}= 4 M_{\substack{intra\\same}}  - 4 M_{\substack{intra\\same}}  + 4 M_{\substack{intra\\diff}} - 4 M_{\substack{intra\\ diff}}=0 ,\\
\nonumber g_{zz}&= 8 M_{\substack{intra\\same}}  -8 M_{\substack{intra\\diff}} ,\\
\nonumber g_{xx}&=g_{yx}=4M_{\substack{inter\\same}}  - 4 M_{\substack{inter\\diff}} ,\\
\nonumber g_{yy}&=g_{xy}=4M_{\substack{inter\\same}}  + 4 M_{\substack{inter\\diff}} ,
 \end{align}
where the first line evaluates to zero because of the property $u_{K^{+}}^*  (\mathbf{r}) =u_{K^{-}}  (\mathbf{r}) $. Furthermore, we  find 
\begin{equation}
M_{\substack{inter\\diff}}=  \sum_{ \mathbf{R} } e^{i(\mathbf{K}^+ - \mathbf{K}^-)\cdot ( \mathbf{R} + \mathbf{R}_{AB}) }\mathcal{M}_{R+R_{AB}} \equiv 0,
\end{equation}
when summing numerically over $\mathbf{R}=c_1\mathbf{a}_1+c_2\mathbf{a}_2$, with non-zero integers $c_1, c_2$, and extrapolating to infinite lattices. The vector  $\mathbf{R}_{AB}$ connects the A and B sublattices on the two different layers. This property restores the symmetry $g_{xx}=g_{yy}=g_{xy}=g_{yx}=:g_{\perp}$. The two remaining nonzero parameters $g_{zz}$ and $g_{\perp}$  read,
\begin{align}
\nonumber g_{zz} &= 8 \mathcal{M}_{ 0} - 8 \mathcal{M}_{R_{AB}} +8 \sum_{ \mathbf{R}\neq0} (  \mathcal{M}_{R} -  \mathcal{M}_{R+R_{AB}} )\\
\nonumber g_{\perp}&=4 \mathcal{M}_{ 0}+ 4  \sum_{ \mathbf{R}\neq0} e^{i(\mathbf{K}^+ - \mathbf{K}^-)\cdot \mathbf{R} } \mathcal{M}_{R},
\end{align}
which, using Z=1 (assuming the inner core electrons of carbon to be fully screened), evaluate to  $g_{zz}=4.04/\epsilon_C$ eVnm$^2$ and $g_{\perp}=-0.191/\epsilon_C$ eVnm$^2$.

\begin{flushleft} 
\begin{table}[t!]
\begin{center}
 small gap, weak interaction\\
%\resizebox{1\hsize}{!}{
\begin{tabular}{|c|c c c| c|}
\hline
&orbital\; & \;spin & valley & $\TPEn_{2P}$\\
\hline
first& & S=0 & $T^{-x}$  & \\
excited& $\TPorbWF^{a}_{ 1}$& S=1 &  $T^{\pm z}$  & $\TPEn_{ 1}^{a}$ \\
state&& S=1 &  $T^{+x}$  & \\
\hline
ground && S=1 &  $T^{-x}$ & $\TPEn_{ 1}^{s}+W(g_{zz}-4g_{\perp})$\\
state &$\TPorbWF^{s}_{ 1}$& S=0 &   $T^{\pm z}$  & $\TPEn_{ 1}^{s} +W g_{zz}$ \\
 && S=0 & $T^{+x}$ & $\TPEn_{ 1}^{s}+W(g_{zz}+4g_{\perp})$\\
\hline
\end{tabular}\\
\vskip10pt
 Small gap, strong interaction\\
\begin{tabular}{|c|c c c| c|} 
\hline
 &orbital\; & \;spin & valley & $\TPEn_{2P}$\\
\hline
first & & S=0 &  $T^{-x}$  & \\
excited & $\TPorbWF^{a}_{ 2}$& S=1 &  $T^{\pm z}$  & $\TPEn_{  2}^{a}$ \\
state& & S=1 &   $T^{+x}$  & \\
\hline
ground&& S=0 &   $T^{-x}$  & \\
state& $\TPorbWF^{a}_{ 1}$& S=1 &  $T^{\pm z}$  & $\TPEn_{ 1}^{a}$ \\
& & S=1 &  $T^{+x}$  & \\
\hline
\end{tabular}\\
\vskip10pt
 large gap, strong interaction\\
\begin{tabular}{|c|c c c| c|}
\hline
&orbital\; & \;spin & valley & $\TPEn_{2P}$\\
\hline
first && S=1 & $T^{-x}$  & $\TPEn_{ 1}^{s}+W(g_{zz}-4g_{\perp})$\\
excited &$\TPorbWF^{s}_{ 1}$& S=0 & $T^{\pm z}$  & $\TPEn_{ 1}^{s} +W g_{zz}$ \\
state&& S=0 &  $T^{+x}$ & $\TPEn_{ 1}^{s}+W(g_{zz}+4g_{\perp})$\\
\hline
ground& & S=0 & $T^{-x}$  & \\
state& $\TPorbWF^{a}_{ 1}$& S=1 & $T^{\pm z}$ & $\TPEn_{ 1}^{a}$ \\
&& S=1 &  $T^{+x}$  & \\
\hline
\end{tabular}
\end{center}
\label{tab:E2PTot}
\caption{States structure and total energies of the \TP states in orbital, spin, and valley space.}
\end{table}%
\end{flushleft} 

\section{Properties of the low-energy two-electron states}

In table I we summarize the orbital, spin, and valley configuration alongside with the total energy of the GS and first excited state of two interacting electrons in the QD, for the different cases shown in \fig 3 of the main text.

\end{document}